\documentclass[format=acmsmall, review=false, screen=true]{acmart}

\usepackage{booktabs} 

\usepackage[ruled]{algorithm2e} 

\SetAlFnt{\small}
\SetAlCapFnt{\small}
\SetAlCapNameFnt{\small}
\SetAlCapHSkip{0pt}
\IncMargin{-\parindent}

\acmJournal{CSUR}
\acmVolume{1}
\acmNumber{0}
\acmArticle{1}
\acmYear{2017}
\acmMonth{0}
\copyrightyear{2017}

\setcopyright{acmlicensed}

\acmDOI{0000001.0000001}

\received{XXX YYYY}
\received[revised]{XXX YYYY}
\received[accepted]{XXX YYYY}

\begin{document}
\title[Interaction Methods for Smart Glasses]{Interaction Methods for Smart Glasses}    
\author{Lik-Hang, Lee}
\author{Pan, Hui}
\orcid{1234-5678-9012-3456}
\affiliation{%
  \institution{The Hong Kong University of Science and Technology}
  \streetaddress{The Hong Kong University of Science and Technology
Clear Water Bay, Kowloon, Hong Kong}
  \city{Hong Kong}
  \state{China}
  \country{USA}}

\begin{abstract}
Since the launch of Google Glass in 2014, smart glasses have mainly been designed to support micro-interactions. The ultimate goal for them to become an augmented reality interface has not yet been attained due to an encumbrance of controls. Augmented reality involves superimposing interactive computer graphics images onto physical objects in the real world. This survey reviews current research issues in the area of human-computer interaction for smart glasses. The survey first studies the smart glasses available in the market and afterwards investigates the interaction methods proposed in the wide body of literature. The interaction methods can be classified into hand-held, touch, and touchless input. This paper mainly focuses on the touch and touchless input. Touch input can be further divided into on-device and on-body, while touchless input can be classified into hands-free and freehand. Next, we summarize the existing research efforts and trends, in which touch and touchless input are evaluated by a total of eight interaction goals. Finally, we discuss several key design challenges and the possibility of multi-modal input for smart glasses. 
\end{abstract}

%
%


%
%


\thanks{

  Author's addresses: L.H. Lee, Department of Computer Science and Engineering, The Hong Kong University of Science and Technology}

\maketitle

\renewcommand{\shortauthors}{Lee and Hui}

\section{Introduction}

In recent years, smart glasses have been released into the market. Smart glasses are equipped with a see-through optical display, which is positioned in the eye-line of human users. The human user can view both the real-world environment and the virtual contents shown in the display, which is regarded as the concept of augmented reality \cite{[110]Poupyrev:2002:DGA:619073.621931}. Currently, augmented reality on mobile devices is dominated by smartphones. For example, one of the biggest smartphone manufacturers, Apple Inc. has launched its augmented reality toolkit, namely ARKit \cite{[115]ARKit}. The shift in mobile devices from smartphones to smart glasses will happen over the next decade. It is projected that smart glasses will become the next leading mobile device after the smartphone, according to market research conducted by Digi-captial \cite{[116]Digi}. Thus, smart glasses have great potential in becoming the major platform for augmented reality. 

According to the figures forecast by Digi-Capital \cite{[116]Digi}, the market value of augmented reality will hit 90 billion US dollar by 2020, in which no less than 45 \% of the market share will be generated by the hardware for augmented reality. In the report by CCS Insight \cite{[117]Insight}, it is estimated that around 14 million of the virtual and augmented reality headsets will be sold by 2020 with a market value of 14.5 billion US dollar. One of the challenges that device manufacturers encounter, before their smart glasses become widespread in the market, is the usability issue. The interaction between human user and smart glasses is still encumbered and problematic. That is, the virtual content on the optical display are not touchable and thus direct manipulation becomes a fatiguing and error-prone task. Additionally, compared with smartphones, smart glasses have more challenging issues such as reduced display size, small input interface, limited computational power, and short battery life \cite{[111]7273236}.

Google Glass \cite{[6]GoogleGlassProject} is the first of its kind in the market. Due to its small form size, only swipe gestures are accessible for the user input and thus the operating system is designed as a series of pixel cards, namely Timeline. Users can swipe over the pixel cards and select the target pixel card. However, this design has potential pitfalls such as limitations in micro-interaction, long search time when pixel card number is large, and so on. Similar to the desktop computer and smartphone, other successors of smart glasses have applied the traditional custom of the WIMP (Windows, Icons, Menus, Pointers) paradigm in their interfaces. However, the default interaction methods available on smart glasses such as touch pad and button inputs are far from satisfactory. The users may find it difficult to accomplish their tasks in the interface under the WIMP paradigm by using these default interaction methods, for instance, the long task completion time, high error rate in item selection, and so on. However, there exists no other standard and mature methods for the interaction between smart glasses and human users.

To tackle this problem, we explore various gestural interaction approaches supported by either the peripheral sensors on additional devices, or embedded sensors in the smart glasses. Gestural input refers to the capturing of the body movements of human users that instructs the smart glasses to execute specific commands. The sensors on the additional devices (e.g. wrist band) or embedded sensors in smart glasses can capture the user's gestures such as drawing a stroke, circle, square, or triangle \cite{[99]Groenewald:2016:UMH:3056355.3056398}. The captured gestures are then converted into input commands according to the gesture library. For example, the possible input commands can be to select a character on the keyboard in the virtual interface shown on the optical display of smart glasses, choosing an app icon on the main menu of the starting page, as well as moving a 3D object from one location to another in the augmented reality environment.

In accordance with the above problem, this survey mainly focuses on research issues related to interaction methods between smart glasses and human users. Equivalently, we focus on the needs of human users in operating the smart glasses. We compare the gestural interaction methods using touch or touchless techniques. We also present the opportunities for using multi-modal methods for the hybrid user interface in augmented reality. In summary, the framework of this survey covers the following areas.

\begin{enumerate}
\item Introduction to Smart Glasses and issues with human-smart glasses interactions (Section 2). Google Glass is the first example of smart glasses in the market, which provides new opportunities for user interaction and the challenges in interaction design to researchers. We evaluate a number of popular smart glasses on the market, and their sensors and the corresponding interaction methods of those smart glasses. 
\item Touch-based interaction approaches (Section 3.1). The user-friendliness of smart glasses is crucial, which becomes an important issue to design easy-to-use and robust interaction techniques. We present various approaches of touchless input to operate smart glasses with external devices or additional sensors.
\item Touchless interaction methods (Section 3.2). Apart from the touch-based techniques with external devices, a number of touchless techniques exist in the literature. We review two primary techniques (Hands-free and Freehand interactions) that enable smart glass users to perform input on smart glasses.
\item Existing research efforts and trends (Section 4). We summarize touch and touchless inputs into four categories and compare them with a total of eight interaction goals. Their potencies and research trends are accordingly discussed.
\item Challenges of interactions on smart glasses (Section 5). All the interaction methods using additional devices or embedded sensors share a common goal. That is, users can perform fast and natural interaction with augmented reality on smart glasses. We present the key challenges of interaction methods to the hybrid user interface in augmented reality. 
\end{enumerate}

\section{Preliminary- the introduction of smart glasses and their sensors}
\label{sec:two}
Smart glasses are head-worn mobile computing devices, which contain multiple sensors, processing capabilities and optical head-mounted displays (OHMDs). With the processing capabilities and the OHMD, the users of smart glasses can view augmented information that is overlaid on the physical world. These capabilities provide great potential to achieve real-time and enriched interaction between the smart glasses user and the physical world with augmented information. Equivalently, the smart glasses wearer can interact with the augmented reality environments. In order to achieve the two-way interactions between the user and the smart glasses, two important requirements should be fulfilled. 

First, smart glasses can provide a clear and stable output on the OHMD to the smart glasses wearer. The smart glasses wearer finds it very difficult to see the content in augmented reality if the output on the OHMD is too small or unclear in some illuminated conditions such as outdoor environments. However, this is highly related to the technical specifications of the smart glasses and thus not the focal point of the survey paper. Second, smart glasses should offer an easy and effortless manner with which to operate them under appropriate ergonomic considerations. The smart glasses user can perform inputs through various actions (e.g. head movement, hand gesture, voice input, etc.) and the sensors embedded into the smart glasses identify the actions of the user. The input of the wearer can be processed into instructions for user interaction with virtual content superimposed onto the physical world. 

This section first includes several significant examples of smart glasses, ranging from the very first prototype (head-worn computer) proposed in the lab to the recent commercial product (smart glasses) available in the market. We can see the advancement of smart glasses has developed from being a bulky and cumbersome backpack to the current lightweight wearables. Next, we depict the sensors available for today's nowadays commercial smart glasses. The usage of sensors for the corresponding input methods will be briefly explained in this section, while the details of the input methods in the wide body of literature is discussed in the next section.

\subsection{Examples of smart glasses}
\begin{figure}
  \includegraphics[width=1\textwidth]{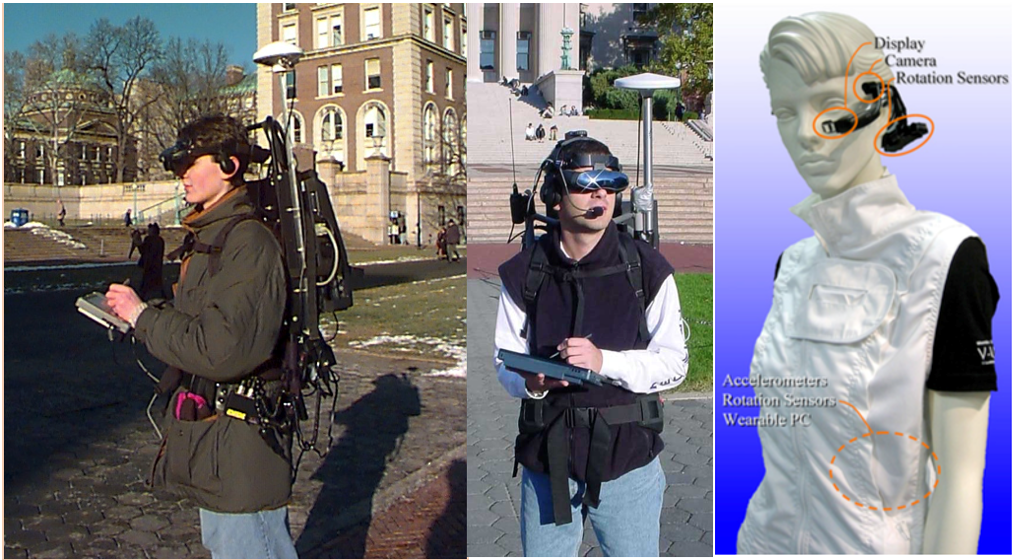}
  \caption{The side view (Left) and front view (Middle) of Touring Machine \cite{[1]Arango:1993:TMS:151233.151239} and Weavy \cite{[2]2002} (Right)}
  \label{fig:Touring and Weavy}
\end{figure}

\paragraph{The Touring Machine}
The first and historically significant example of smart glasses can be traced back to the Touring Machine \cite{[1]Arango:1993:TMS:151233.151239}, which was proposed in 1997 by Feiner et al. The Touring Machine is a prototype machine designed for urban exploration. In the demonstration, it is used for the navigation of the campus area. The machine consists of a wearable see-through display with built-in orientation detector, a stylus and a trackpad on a handheld computer, a GPS receiver, peripherals for the internet connection, and a desktop computer stowed in the backpack of the user. Figure \ref{fig:Touring and Weavy} shows the appearance of the Touring Machine. 

In the campus navigation demonstration, through the see-through display, the user can see information about the surrounding buildings in the campus (for instance, names of buildings and corresponding departments) as well as a number of choices on a virtual menu such as finding the user location, showing the department information, removal of digital overlay, and so on. The user can access the digital overlaid menu with the trackpad. Also, the orientation detector guides the users to the orientation of the destination building. The compass system presents a compass pointer on the see-through display. The color of the points will change from green to red if the user deviates from the target building more than 90 degrees. 

\begin{figure}
  \includegraphics[width=1\textwidth]{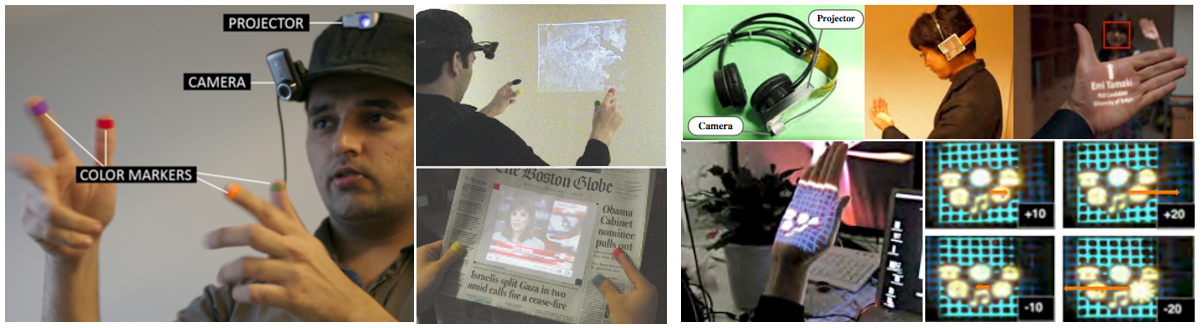}
  \caption{WUV \cite{[3]Mistry:2009:WWU:1520340.1520626} (Left) and BrainyHand \cite{[4]Tamaki:2010:BWC:1842993.1843070} (Right)}
  \label{fig:WUV and BrainyHand}
\end{figure}

\begin{figure}
  \includegraphics[width=0.6\textwidth]{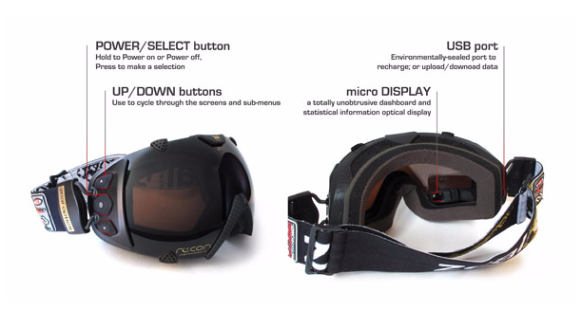}
  \caption{Transcend HUD\cite{[5]Recon}}
  \label{fig:Transcend_HUD}
\end{figure}

\begin{figure}
  \includegraphics[width=1\textwidth]{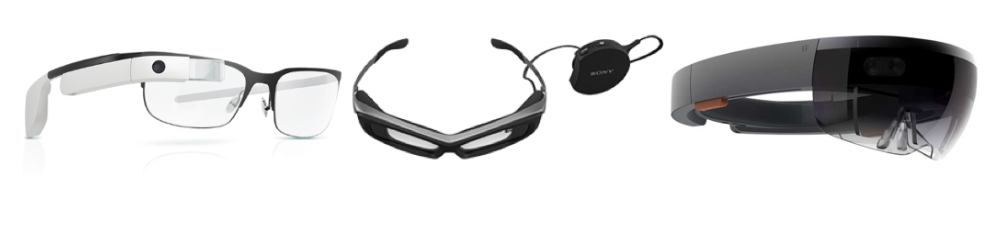}
  \caption{Google Glass \cite{[6]GoogleGlassProject} (Left), Sony SmartEyeGlass \cite{[7]SonySmartEyeglasses} (Middle),
 Microsoft Hololens \cite{[8]MicrosoftHololens} (Right)}
  \label{fig:three smart glasses}
\end{figure}

Even though the system is cumbersome and heavyweight, in comparison to today's smart glasses, it is a well-defined example of the early development of augmented reality on mobile devices, where the features of Touring Machine are driven by a GPS. This is basically the same as today's GPS-driven mobile applications. Also, it presents a rudimentary approach to the interaction with a digital overlaid menu in augmented reality by using trackpad and stylus. 
\paragraph{Weavy}
Weavy \cite{[2]2002} is a lightweight head-worn mobile wearable, which is comprised of a single-eyed head-mounted display with the capabilities of wireless connection (Figure \ref{fig:Touring and Weavy}). All the frames captured from the camera on the device are transmitted to the back-end server that handles the offloading of computer-vision tasks. Compared with the Touring Machine, Weavy demonstrates a working prototype, which is closest to today's smart glasses. However, due to the limitations of computing power in 2002, the image frames are processed in the back-end server. 
\paragraph{WUV and BrainyHand}
WUV \cite{[3]Mistry:2009:WWU:1520340.1520626} and BrainyHand \cite{[4]Tamaki:2010:BWC:1842993.1843070} can be regarded as a variation of Weavy. These wearables, as shown in Figure \ref{fig:WUV and BrainyHand}, have a similar purpose to smart glasses but the key difference is that a laser projector substitutes the optical output on the head-mounted display. The wearables can show the augmented information onto the user's palm or nearby surface such as an interior wall and newspaper. As these devices are miniature in size, trackpads or buttons are not available for user input. The user input of these wearables are mainly supported by hand gestures. In WUV, the wearers have to stick colorful markers on their hand for the recognition of the user's hand gesture input, while BrainyHand is able to detect simple hand gestures, such as zooming in and out, by calculating the distance between the skin surface of the user's hand and the head-mounted camera.
\paragraph{Transcend HUD}
Before the commencement of Google Glass, Transcend \cite{[5]Recon} is the first example of commercial smart glasses launched in 2010. They are ski-goggles that are equipped with a Heads-Up Display (Figure \ref{fig:Transcend_HUD}). The data is displayed on a small screen on the outer edge of a skier's peripheral vision. With the assistance of location-aware features driven by a built-in GPS, the smart glasses can notify a skier about the real-time performance such as speed, elevation, airtime and navigation. 
\paragraph{Google Glass and Sony SmartEyeGlass}
Google Glass \cite{[6]GoogleGlassProject}, which was released in 2014, is a light-weighted and self-contained head-mounted computer with a set of sensors such as accelerometer, gyroscope and magnetometer (Figure \ref{fig:three smart glasses}). Compared with the previous example, the virtual content is visible in a see-through optical display, which is made of liquid crystal on silicon (LCoS) with LED illumination. Thus, the smart glasses are capable of superimposing virtual content such as text and images onto the user's field of view (FOV). It allows the wearer to perform micro-interaction with the smart glasses such as map navigation, photo or video capturing, and receiving notification/message. Regarding the input method, voice command (speech recognition) is the major method of operating Google Glass. Similar to Weavy, the task of natural language processing is offloaded to Google's cloud server that analyze the user's input, due to the limited computational capabilities of Google Glass. Epson \cite{[7]SonySmartEyeglasses} released its first smart glasses in 2015, as shown in  (Figure \ref{fig:three smart glasses}). It is a similar product to Google Glass with a considerably larger optical display. Also, its input method relies on a touch-sensitive external controller that enables the user to operate a mouse cursor in the WIMP interface.

\paragraph{Microsoft Hololens}
Microsoft Hololens \cite{[8]MicrosoftHololens} are recently launched smart glasses that are equipped with powerful computer chipsets and a state-of-the-art display with wider FOV than the aforementioned examples  (Figure \ref{fig:three smart glasses}). The chipsets create a more immersive environment, which allows the user to pin holograms onto the surrounding physical environment. The holograms can be represented in the taxonomy of a 2D interface and a 3D object. 2D objects can be virtual windows/menus, writing notes, gallery, video, while 3D objects can be sphere, cube, animal, planet, etc. Also, it supports multi-modal input including a head gesture for cursor movement, two simple hand gestures (tap and blooming), and voice command. Although they are the most powerful self-contained smart glasses on the market, they are considered as obtrusive with a bulky design and lack of mobility in outdoor environments. It is suggested that a wearable interface must be ready for mobility or in-situ use \cite{[25]Ens:2016:CRI:2983310.2985757}.
\paragraph{Summary}
Today's smart glasses are regarded as the beginning of augmented reality on mobile devices. However, they are considered as a rudimentary product because major constraints, such as weak processors, short battery life, small screen size, have not yet been solved. Considering the focal point of this paper, the input methods for smart glasses are not well-defined. Even though the projection of smart glasses is promising, we are not clear if smart glasses will be adopted by users for daily usage in the same way as today's smartphones, as the issues of battery life and input methods are problematic. However, it seems that smart glasses will first serve as some specialized task-oriented devices, for instance, industrial glasses, smart-helmets, sport-activity coaching devices, and the like. \cite{[109]RAUSCHNABEL2015635}. 

\subsection{Sensors on the smart glasses and the input methods}

\begin{figure}
  \includegraphics[width=1\textwidth]{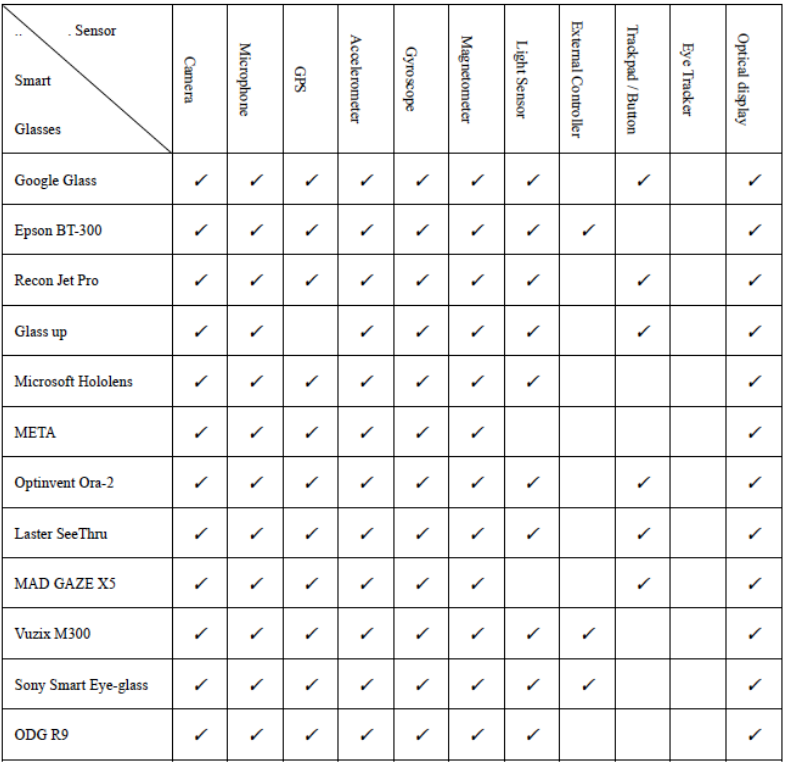}
  \caption{Sensors on Commercial Smart Glasses}
  \label{fig:Sensor Table}
\end{figure}

Figure \ref{fig:Sensor Table} shows the sensors on the smart glasses available in the market that support various input methods in the practice and the literature. The only exception is the optical display in the final column, which is the standard component for optical output. The sensors are briefly explained as the following.
.

\begin{figure}
  \includegraphics[width=1\textwidth]{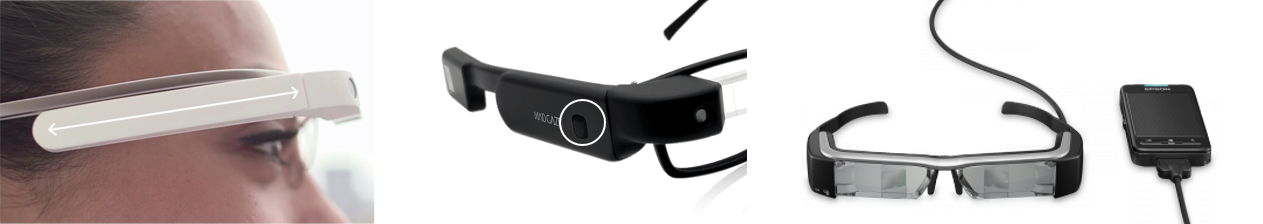}
  \caption{Trackpad on spectacle frame (Left), Button (Middle), External controller wired with Espon smart glasses (Right)}
  \label{fig:Button_Touchpad_controller}
\end{figure}

\paragraph{Camera}
It is an optical instrument for recording or capturing images, which may be individual still photographs or sequences of images constituting videos or movies \cite{[9]Wikipedia:Camera}. Camera are one of the standard components on smart glasses. Among the available smart glasses, the majority of them (9 out of 12) are equipped with RGB camera that is only designed for monocular vision. This is mainly restricted by the requirements of the product size, as depth cameras and infrared cameras are bulky and heavyweight. Therefore, we find that the remaining three smart glasses support depth measurement and infrared data, in which Microsoft Hololens and ODG have depth cameras, and META supports infrared vision. The cameras on the glasses can support various computer-vision tasks and their capabilities are subject to the types of camera. When the camera comes to the domain of user input, it is usually for capturing a wearer gesture, in particular of hand gestures.
\paragraph{Microphone}
It is a transducer that converts sound into an electrical signal \cite{[10]Wikipedia:Microphone}. The electrical signal can be further processed by speech recognition. The recognized speech is used for input to the smart glasses. All smart glasses have microphones embedded into their circuit board. This implies that today's smart glasses support voice input from users. One of the reasons is that the recent advancements in speech recognition makes voice input accurate and responsive.
\paragraph{Global Positioning System(GPS)}
It is a global navigation satellite system that provides geo-location and time information to a GPS receiver anywhere on the Earth \cite{[11]Wikipedia:GPS}. The GPS enables the smart glasses to support various geo-location based applications. For instance, the smart glasses can tell the users about the current position of the wearer, or driving directions. From the results, 11 out of 12 smart glasses have GPS, which makes them ready for GPS based augmented reality applications.
\paragraph{Accelerometer}
The sensor is designed for measuring proper acceleration, which is defined as the rate of change of velocity of a body in its own instantaneous rest frame \cite{[12]Wikipedia:Accelerometer}. All smart glasses have accelerometer. Smart glasses can measure the acceleration force along the x, y, and z axis, as well as gravity force. This allows the smart glasses to record the motion input from the wearer, for instance, understanding the status and activities of the wearer like being stationary, walking, running, and so on. In addition, knowing the status and activities of the user can help in designing user input in a more precise and subtle manner \cite{[13]Tung:2016:EHI:2906388.2906394}. For example, the wearer performs head gestural input to the smart glasses but the accuracy of the gesture recognition can be influenced by the other simultaneous motions, for instance, the walking status of the wearer. Thus, the unwanted motion from walking can be alleviated by the measures taken by the accelerometer. 
\paragraph{Gyroscope}
The sensor is an infrastructure which measures the orientation of the wearer on the basis of the principle of angular momentum \cite{[14]Wikipedia:Gyroscope}. The rate of rotation around the x, y, and z axis are measured by the infrastructure. Identical to the accelerometer, the gyroscope exists in all smart glasses as gyroscopes and accelerometers are commonly integrated into today's manufacturing standard. Regarding the input approach of smart glasses, a gyroscope can measure the angular velocity of the wearer's head. Therefore, smart glasses can measure the head movement of the wearer and hence support head gestural input.
\paragraph{Magnetometer}
The sensor is an instrument that measures the strength and direction of magnetic fields \cite{[15]Wikipedia:Magnetometer}. Many smartphones have magnetometers and they serve as compasses in various mobile application especially for navigation and maps. Similarly, all smart glasses have magnetometers as they have inherited the requirements for mobile applications on smartphones. It is projected that smart glasses have the same potential to measure the wearer's mobility and perform various mobile applications as appear in today's smartphone when both the accelerometer and gyroscope are considered. 
\paragraph{Light sensor}
The sensor is a detector of light or other electromagnetic energy \cite{[16]Wikipedia:Photodetector}. As for the smartphone, the touchscreen display adjusts its brightness subject to the ambient light. Likewise, the optical display of smart glasses adjusts the brightness if the ambient light affects the readability of content. Thus, the light sensor provides smart glasses the capability of  automatically adjusting the brightness of the display in various light conditions. From the results, only two of the surveyed smart glasses are not equipped with light sensors. The META is currently designed for Augmented reality in indoor environments. However, Mad Gaze X5 is designed for both indoor and outdoor environments. The lack of a light sensor impacts on the readability of content in outdoor environments. 
\paragraph{Tangible interface}
This category refers to the use of an external controller, trackpad and button, which allows the wearers to interact with the digital interface of the optical display of smart glasses (Figure \ref{fig:Button_Touchpad_controller}). The external controller provides a more efficient and easier control than the trackpad and button located on the body of the smart glasses. However, the external controller is cumbersome if the wearer's hands are occupied and thus it is not convenient for the wearer to perform other tasks simultaneously in augmented reality. The trackpad and button have no issue with the above problem but the operations on the small touch surface of the button and trackpad of the smart glasses causes two major problems \cite{[23]Chew:2010:UEU:1868914.1868930}. First, muscle fatigue is the main issue as wearers need to raise their hands to touch the button and trackpad. Prolonged use is not favorable for the wearer \cite{[100]Hincapie-Ramos:2014:CEW:2598784.2602795}. Second, the small surface of the button and trackpad requires subtle finger movements and therefore deteriorates the task performance. Nevertheless, the above input methods are commonly used for smart glasses. Out of 12 smart glasses 6 provide buttons and trackpads for manipulating the items and objects on the smart glasses interface, while 3 out of 12 smart glasses have external controllers that enable users to control a cursor on the smart glasses interface. The remaining three smart glasses rely on the gestural input supported by various types of cameras. META utilizes an infrared camera to detect hand gestural inputs from the wearers, and Microsoft Hololens and ODG R9 utilize depth camera to capture hand gestural input. The results show that the hand gestural input is an alternative to the tangible interface, because of its advantages such as intuitiveness and naturalism \cite{[17]Norman:2010:GIS:1836216.1836228}. 

\paragraph{Eye tracker}
It is a device for measuring eye positions and eye movement. Nowadays, it is mainly applied in the virtual reality such as FOVE \cite{[18]FOVE}. Unfortunately, none of the smart glasses supports the eye tracking function. Here is an example showing the potential of using a multi-modal input approach of eye-tracking and physical interface. When only the tangible interface is accessible, it is difficult for users to select one small object in cluttered environments. The eye-tracking technology can be used to quickly spot and locate the object that the user intends to select, which is driven by eye movement \cite{[22]Toyama:2012:GGO:2168556.2168570}. Afterwards, the user can manipulate the object by the tangible interface such as button or external controller. Alternatively, the combination of eye tracking and hand gesture can achieve object location and selection \cite{[19]Slambekova:2012:GGB:2407336.2407380}.  
\paragraph{Summary}
To conclude, today's smart glasses have evolved from a bulky and heavy machine located in the user's backpack  to lightweight wearables. The ways of showing virtual content are unified to see-through optical displays from head-mounted displays and projections onto nearby surfaces. The twelve surveyed smart glasses are equipped with cameras, microphones, accelerometers, gyroscopes, magnetometers, which are widely available in many smartphones. GPSs and light sensors are important in aiding smart glasses to adapt with the mobile applications in outdoor environments. Eye tracker is gaining popularity in the field of virtual reality, but none of the smart glasses manufacturers have taken eye tracker into their commercial products and the technology of eye tracker for augmented reality smart glasses is in its infancy, even though a few lower cost add-on components for eye tracking on head-worn computers have been proposed \cite{[20]Stengel:2015:ASB:2733373.2806265,[21]Shimizu:2016:EMI:2968219.2968274}. Not surprisingly, the kind of hand gestural interaction has first been applied to the commercial products from research. Many approaches have been widely proposed in the literature, ranging from head gestures, gaze interaction, to touch interface on different parts of the human body.  In the next section, we investigate various interaction approaches for smart glasses in the literature, which are supported by the embedded sensors introduced in this section and other additional sensors.

\section{Interaction approaches for smart glasses}
\label{sec:three}
Nowadays, touchscreen input is the primary interaction modality for today's smart devices, and these touchscreens are sized from smart wristbands to smartphones. As for the smart wearables, such as smart glasses, speech recognition is the major input of choice because these wearable devices do not have a touch-screen display that serves as the input device. Despite the fact that touch screens are popular in smartphones and smart watches, the screen touch interfaces have not moved into small-sized smart devices with following reasons \cite{[26]Colaco:2013:MCL:2501988.2502042}. A touch screen interface does not fully take advantage of human dexterity. It requires the user to touch a small screen on the device repetitively and constantly, and hence touching the screen for input occludes the user's sight of the display. This makes the simple tasks like menu navigation becoming repetitive and tedious actions. Therefore, studies in the literature have proposed numerous approaches to interact with smart wearables of small size including smart glasses. Offering smart glasses with better input approaches makes the interaction experience more intuitive and efficient, which enables the users to handle more complicated and visually demanding tasks. In other words, the enhanced interaction experience brings smart glasses from their limited usage of micro-interactions to daily usage as seen in today's smartphones. In this section, we focus solely on the interaction approaches for smart glasses.

\begin{figure}
  \includegraphics[width=1\textwidth]{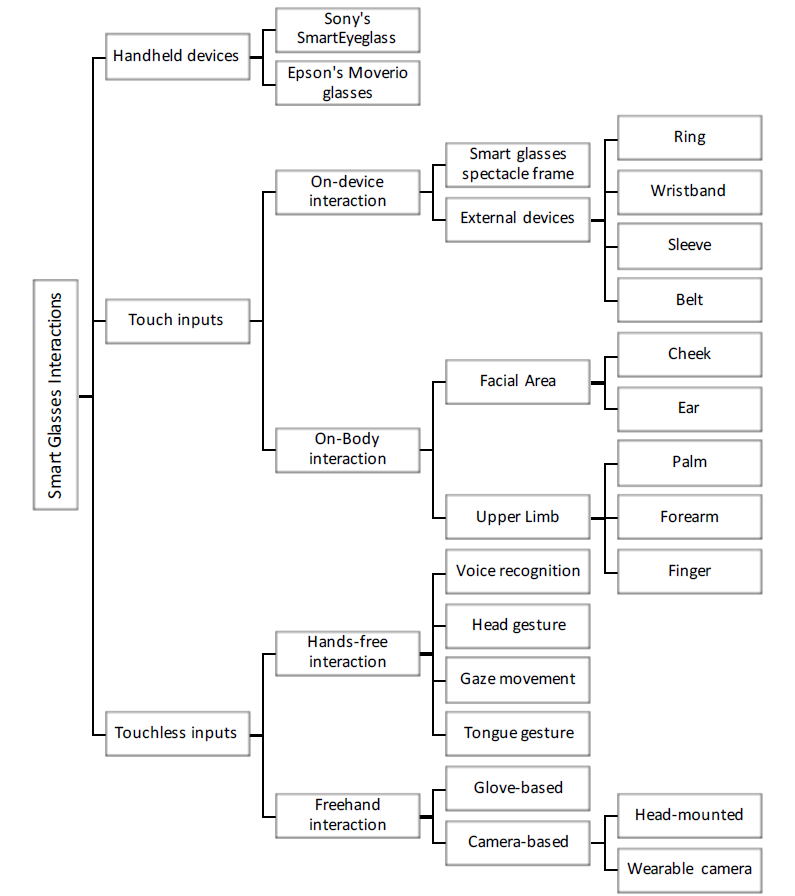}
  \caption{Classification of interaction approaches for smart glasses}
  \label{fig:classification_scheme}
\end{figure}

There are multiple dimensions for classifying interaction approaches, for instance, Vision-based and Non-vision based, Gesture-based and Non-gesture based \cite{[27]Bertarini}. An alternative dimension is to divide the interaction approaches into 3 classes, which are handheld, touch, and touchless \cite{[24]Tung:2015:UGI:2702123.2702214}. First, handheld refers to the input type that makes use of handheld controllers, such as smartphones, and the wired trackpads linked with Sony's SmartEyeglass and Epson's Moverio glasses. Second, touch refers to non-handheld touch input, such as gestures and tapping on body surfaces, touch-sensing wearable devices (e.g. smart rings, smart wrist band, watches, and spectacle frame of smart glasses), as well as touch interface on the user's body. This class is characterized by the presence of tactile feedback. Third, touchless refers to non-handheld, non-touch input, such as mid-air hand gestures, head and body movements, gaze interaction, and voice recognition. In contrast with the second class, this class does not involve tactile feedback from touch but tactile feedback can be augmented by devices (e.g. haptic feedback from a haptic glove \cite{[96]Hsieh2014} or a head-worn computer \cite{[92]Kangas:2014:GGH:2556288.2557040}). The first class have been briefly explained with the tangible interface in Section 2. The remainder of the classes (Touch and touchless) are discussed in this section.  Figure \ref{fig:classification_scheme} depicts the classification of interaction approaches  proposed in this survey.

\subsection{Touch inputs}

\subsubsection{On-device interaction}
\begin{figure}
  \includegraphics[width=1\textwidth]{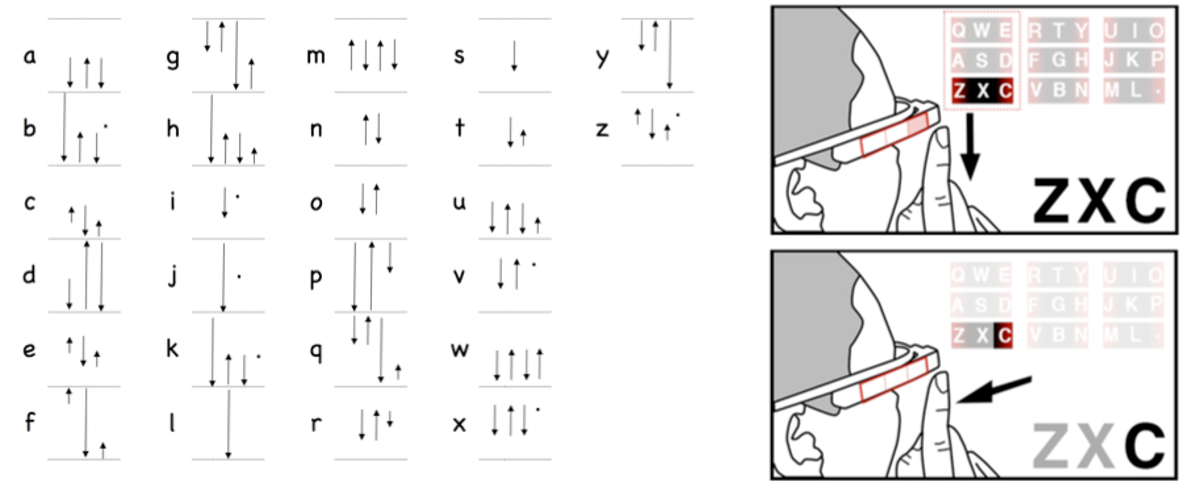}
  \caption{Two dimensional uni-stroke writing system \cite{[65]Yu:2016:OHI:2858036.2858542} (Left) and SwipeZone \cite{[66]Grossman:2015:TGA:2785830.2785867} (Right)}
  \label{fig:Writing System}
\end{figure}

\begin{figure}
  \includegraphics[width=1\textwidth]{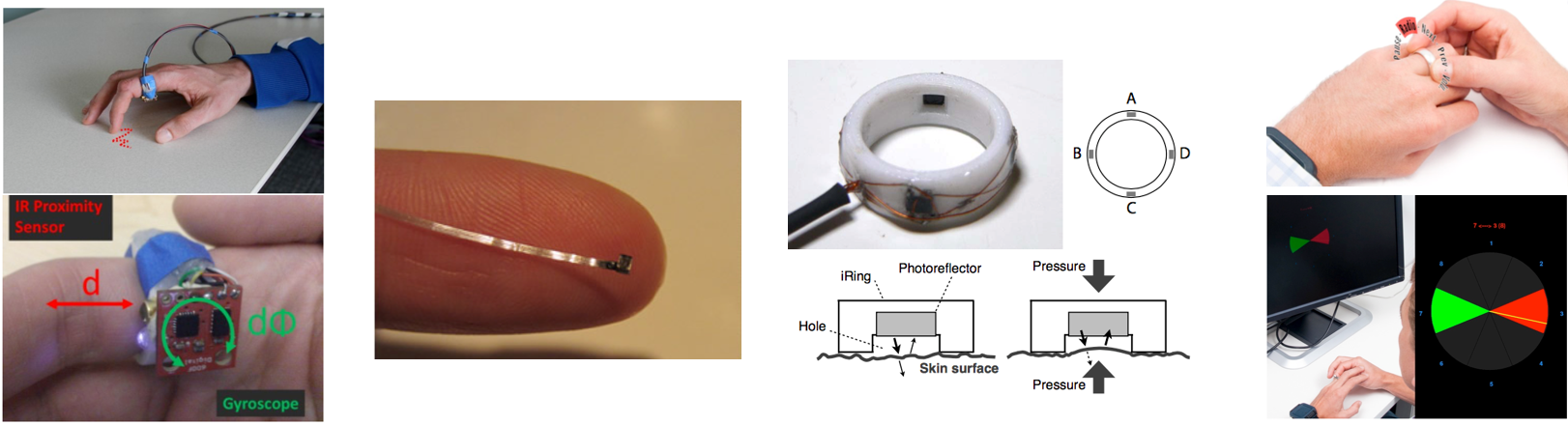}
  \caption{LightRing \cite{[60]Kienzle:2014:LAI:2642918.2647376} (Left), MagicFinger \cite{[61]Yang:2012:MFA:2380116.2380137} (Left Middle), iRing \cite{[62]Ogata:2012:IIR:2380116.2380135} (Right Middle), Nenya \cite{[63]Ashbrook:2011:NSE:1978942.1979238} (Right)}
  \label{fig:Ring}
\end{figure}

On-device interaction means the users can perform gestural input on a sensible surface of various devices such as the body of smart glasses and peripheral sensors on external devices, which serves as an augmented touch surface for user inputs. 
\paragraph{Touch interface on smart glasses}
Google glasses have a touchable spectacle frame, where a swipe gesture can be acted on the frame. Researchers propose swipe-based gesture for text entry \cite{[65]Yu:2016:OHI:2858036.2858542,[66]Grossman:2015:TGA:2785830.2785867}. In Yu et al's work \cite{[65]Yu:2016:OHI:2858036.2858542}, an unistroke gesture system is proposed (Figure \ref{fig:Writing System}). Each character is represented by a set of two dimensional uni-strokes. These stroke sets are designed for easy memorization. For example, the character `a' is comprised of three swipes of 'down-up-down' that mimics the stroke of handwriting. In SwipeZone \cite{[66]Grossman:2015:TGA:2785830.2785867}, the touchable spectacle frame on Google Glass are divided into three zones (back, middle and front). A character can be quickly chosen by two swipes on these zones (Figure \ref{fig:Writing System}). The first swipe selects the character block consisting of 3 characters. The second swipe chooses the target character inside the block. On the other hand, other works focus on the optimal use of the external controller wired with smart glasses to achieve faster text entry. The external controller allows users to operate the pointing device, that is, the cursor, and select keys on a virtual on-screen keyboard. Various arrangements of text input interface are considered in the literature such as Dasher \cite{[67]Ward:2000:DDE:354401.354427}, as well as AZERTY and QWERTY keyboards \cite{[64]7170665}. 
.
\paragraph{Physical forms of external devices}
As smart glasses have a reduced form size and weight, the need for complementary interaction methods are evolving. External devices can be made in various physical forms such as rings \cite{[60]Kienzle:2014:LAI:2642918.2647376,[61]Yang:2012:MFA:2380116.2380137,[62]Ogata:2012:IIR:2380116.2380135,[63]Ashbrook:2011:NSE:1978942.1979238}, wristbands \cite{[56]Rekimoto:2001:GGU:580581.856565,[57]Ham2014}, sleeves \cite{[52]Schneegass:2016:GUT:2971763.2971797}, and belts \cite{[53]Dobbelstein:2015:BUT:2702123.2702450}. Instrumental glove is excluded from this category because of its purpose for mid-air interaction \cite{[55]4539650}. The on-device interactions are precise and responsive. That is, the spatial mapping between the sensible interface on external devices and the smart glasses' virtual interface allows accurate input and fast repetition. However, the major drawback is the existence of the device itself and the time required for putting on the device \cite{[25]Ens:2016:CRI:2983310.2985757}.
\paragraph{Finger-worn device}
Finger-worn devices (Figure \ref{fig:Ring}) have gained a lot of attention in recent years, as these devices encourages small, discreet, and single-handed movements \cite{[58]Shilkrot:2015:DDC:2830539.2828993}. LightRing \cite{[60]Kienzle:2014:LAI:2642918.2647376} consists of a gyroscope and an infrared emitter positioned on the second phalanx of the index finger, while MagicFinger \cite{[61]Yang:2012:MFA:2380116.2380137} has an optical sensor positioned on the fingertips. These types of hardware enable stroke-based gestures on any surface. In LightRing, the infrared emitter and gyroscope detect changes in distance and orientation that constitute trajectories on touch surfaces. The miniature optical sensor on Magic Finger detects the direct touch of fingertips on any solid surface. In contrast, iRing \cite{[62]Ogata:2012:IIR:2380116.2380135} and Nenya \cite{[63]Ashbrook:2011:NSE:1978942.1979238} provides a touch surface on the ring. Users can touch these ring surfaces for pointing and flipping gestures. In addition, iRing can detect both the touch on the ring surface and the bending of the finger muscle, in which the gesture combination is enriched. The photoreflector in the ring can detect the changes in pressures from touch and finger bending. Nenya has a magnetometer in the baselet sensing the absolute orientation of finger touch. Ens et al. \cite{[25]Ens:2016:CRI:2983310.2985757} and Nirjon et al. \cite{[121]Nirjon:2015:TWR:2742647.2742665} attempt to further extend the capability of ring-form devices. As ring-form devices own a relatively small sensitive surface, its usage is commonly proposed for tap and swipe gestures. Nirjon et al. \cite{[121]Nirjon:2015:TWR:2742647.2742665} propose a finger-worn text entry system for a virtual QWERTY keyboard. The keys on a QWERTY keyboard are divided into multiple zones in which every zone contains a sequence of 3 consecutive keys. Two steps are compulsory for choosing a key, as follows. In the first step, users select the target zone by moving the hand horizontally and vertically on a surface. Next the user locates the target key by finger movement, as the ring mounted on the middle finger can detect the user's finger movements (middle, index, and ring fingers). Another ring proposed by Ens et al. \cite{[25]Ens:2016:CRI:2983310.2985757} contains an inertia measurement unit and touch surface. This hardware configuration supports tap and swipe gestures during hand gestural input. A depth camera mounted on the smart glasses detects the hand gestures for fast and coarse selection of a window. The user can use a fingertip to point on a virtual object and afterwards interact with the chosen object through the tap and swipe gestures powered by the ring.
.

\begin{figure}
  \includegraphics[width=1\textwidth]{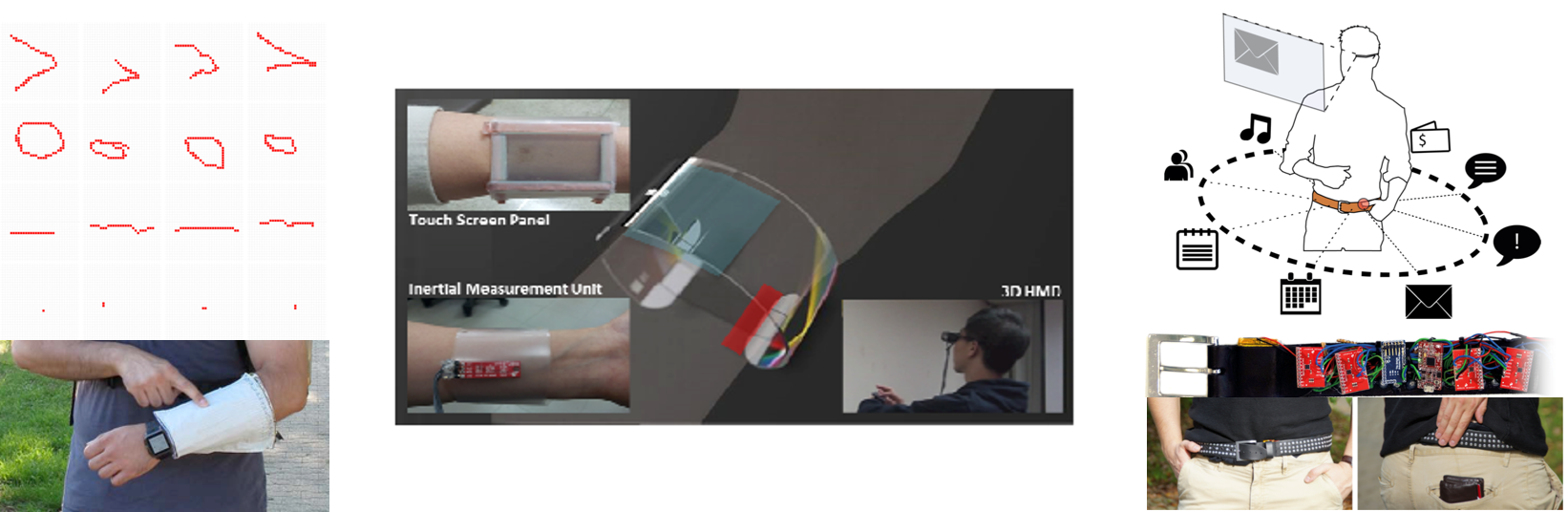}
  \caption{GestureSleeve (Left) \cite{[52]Schneegass:2016:GUT:2971763.2971797}, Smart Wristband \cite{[57]Ham2014} (Middle), Belt \cite{[53]Dobbelstein:2015:BUT:2702123.2702450} (Right)}
  \label{fig:Wrist and Belt}
\end{figure}

\paragraph{Arm-worn device}
These devices have a relatively larger surface than finger-worn devices. Instead, the touch surface is located on the wristband (Figure \ref{fig:Wrist and Belt}). Muscle tension \cite{[56]Rekimoto:2001:GGU:580581.856565} and arm movement \cite{[57]Ham2014} (e.g. wrist rotation) are detected by capacitive sensors and an inertial measurement unit (IMU), respectively. Gesture Sleeve \cite{[52]Schneegass:2016:GUT:2971763.2971797} is a variation of wristband covering the entire area of the forearm with a touch-enabled textile that supports tap and stroke based gestures.
\paragraph{Touch-belt device}
Dobblelstein et al. \cite{[53]Dobbelstein:2015:BUT:2702123.2702450} have proposed a touch-sensitive belt for smart glasses inputs. The belt-shape prototype intends to provide users a larger input surface than the spectacle frame on Google Glass. The touch-sensitive area on the belt (Red circuit boards as shown in Figure \ref{fig:Wrist and Belt}) supports swipe gestures to manipulate the pixel cards on the optical display. The approach is claimed to be unobtrusive as the user do not need to lift the arm and only subtle interaction with the belt is involved. However, this work only considers swipe gestures for Google Glass, while the pointing technique in WIMP paradigm \cite{[54]Jacob:2008:RIF:1357054.1357089} is neglected.
.

\subsubsection{On-body interaction}
Many researches have utilized human skin as the interaction surface. The prominent feature of on-body interaction is to leverage human proprioception as an additional feedback mechanism. That is, a human user can sense the tactile cue when interaction is exerted on the skin's surface. Due to the existence of the tactile cue, on-body interaction has higher performance than touchless input especially mid-air input. Users no longer rely on visual clues to accomplish their tasks when the tactile clue can help them to locate their touch \cite{[28]Gustafson:2013:UPI:2470654.2466114}. In other words, the tactile cue can release the visual attention and achieve eyes-free input that is useful in actions with lower cognitive/physical efforts or lack-of-attention scenarios \cite{[38]Yi:2012:EUM:2207676.2208678}. For instance, when users are walking (i.e., in mobile scenarios), eyes-free input through on-body interaction allows them to pay attention to the surroundings without high attention on the input interface, which could reduce distraction and danger \cite{[44]Fuentes2010}. Besides, users can immerse themselves in augmented reality without switching their attention between the input interface and the virtual contents on the optical display of smart glasses.

A recent work by Wagner et al. \cite{[39]Wagner:2013:BDS:2470654.2466170} investigates the body-centric design space to understand the multi-surface and on-body interactions. Three guidelines for designing on-body interactions are proposed accordingly. Task difficulty, body balance, and interaction effects should be considered together for the on-body interaction. Particularly, the on-body interaction should be selected on stable body parts, such as upper limbs, especially when tasks require precise or highly coordinated movements. In another study conducted by Wagner et al. \cite{[43]Weigel:2014:MTU:2556288.2557239}, the on-skin input on various positions of the upper limbs are studied thoroughly. The user preference shows that the forearm is the highest perceived ease and comfort location (50\%), followed by the back of the hand (18.9), the palm (17.8\%), the finger (7.3\%) and others (6\%). However, the above studies have not considered touch on the facial area. Facial touch has high potential because smart glasses are positioned on the user's head, and at the same time facial touch is proximate to the smart glasses, which serves as an extension of the touch interface on smart glasses, in addition to the benefits such as intuitive and natural interactions \cite{[35]Mahmoud:2011:IHG:2062850.2062879}.

The prior work of on-body interaction have proposed various parts of the human body, such as the palm \cite{[28]Gustafson:2013:UPI:2470654.2466114,[32]Harrison:2011:OWM:2047196.2047255,[33]Harrison:2011:SAS:1978542.1978564,[40]Weigel:2015:IFS:2702123.2702391,[42]Wang:2015:PUP:2785830.2785886,[45]Wang:2015:PUP:2785830.2785885}, the forearm (combined with the back of the hand) \cite{[34]Azai:2017:SMM:3027063.3052959,[34]Azai:2017:SMM:3027063.3052959,[46]Ogata:2013:SAS:2501988.2502039,[47]Lin:2011:PPU:2047196.2047259}, the finger \cite{[37]Huang:2016:DDT:2858036.2858483,[40]Weigel:2015:IFS:2702123.2702391,[41]Yoon:2015:TFT:2677199.2680560}, the face \cite{[29]Serrano:2014:EUH:2611247.2556984}, the ear\cite{[30]Lissermann:2013:EAB:2468356.2468592} for touch input, as the following.
\paragraph{Palm as surface}
The projection-based techniques are first adaptable to smart glasses. OmniTouch \cite{[32]Harrison:2011:OWM:2047196.2047255} is a shoulder-worn wearable proof-of-concept system equipped with depth-sensor and projector. Users can perform multi-touch interaction on their own bodies including the palm. In addition, the projection of virtual contents can be applied to any flat surface. The user can receive tactile feedback from the finger when active touch \cite{[28]Gustafson:2013:UPI:2470654.2466114} is acted on these surfaces. Skinput \cite{[33]Harrison:2011:SAS:1978542.1978564} is an arm-worn wearable hardware with projector and vibration sensors. Instead of using a depth sensor to detect a touching event on an user's skin, an array of tuned mechanical vibration sensors are used to capture wave propagation along the arm's skeletal structure when a finger presses on the skin. 

\begin{figure}
  \includegraphics[width=1\textwidth]{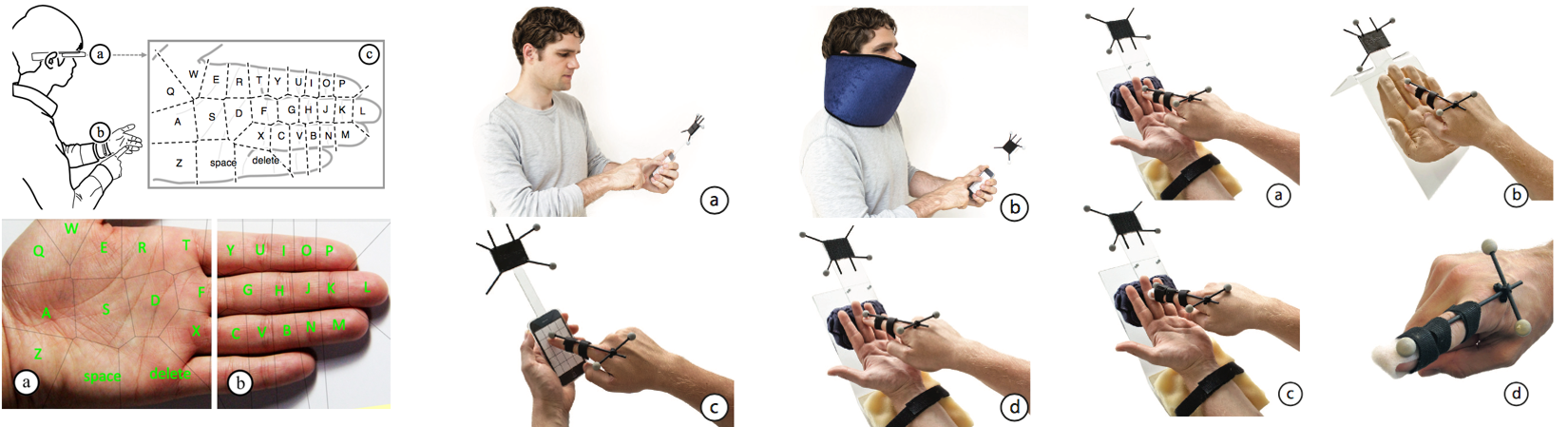}
  \caption{PalmType \cite{[42]Wang:2015:PUP:2785830.2785886} (Left), Gustafson et. al's \cite{[28]Gustafson:2013:UPI:2470654.2466114} 1st experiment (Middle)  and 2nd experiment (Right)}
  \label{fig:Palm}
\end{figure}

PalmType \cite{[42]Wang:2015:PUP:2785830.2785886} is a palm-based keyboard for text entry. Instead of using the projection proposed in OmniTouch \cite{[32]Harrison:2011:OWM:2047196.2047255}, a virtual QWERTY keyboard appears on the optical display of smart glasses (Figure \ref{fig:Palm}). A number of infrared sensors located on the wrist of the user's forearm detect the touch acting on the palm keyboard. Three types of text entry methods are compared in the evaluation - Touchpad on the external controller wired with Epson Moverio glasses, Squared QWERTY keyboard on the palm, and optimized QWERTY keyboard that matches with shape of the user's palm. The results show that PalmType with optimized layout achieved 10 words per minutes, which was 41\% faster than touch pad, and 29\% faster than PalmType with a squared layout. The above results give a cue that the mapping of virtual interfaces on the body surface can influence the task performance. The palm should be treated not only as a writing board surface but also a dynamic interface on the body surface. 

In the above examples, visual clues exist in the form of image projection or virtual images on the palm. The surprising fact is that visual feedback is optional to palm-based interaction. Gustafson et al. \cite{[28]Gustafson:2013:UPI:2470654.2466114} investigate the possibility of palm-based imaginary interfaces. That is, no visual cue appears on the user's palm. Alternatively, an audio system announces instructions to the user rubbing across their palms. In the studies, two experiments have been conducted. According to the first experiment, palm-based imaginary interfaces allow people to interact effectively on the palm without visual feedback. Audio instructions assist users rubbing across their palms. In the second experiment, most of the participants agreed that the tactile sensing on the palm is more important than the tactile cue on the pointing finger. In other words, users rely on the tactile sense on the palm to orient themselves to the targeted item in the imaginary interface. 

A brief description of the two experiments are as follows. Four scenarios are designed in the first experiment  (Figure \ref{fig:Palm}). 1) Palms or a fake phone are in sight, 2) Blindfolded, blocks the sight of participants of their hands, 3) a fake phone surface where a grid is drawn on the surface to guide the participants to find the target item, and 4) Palm.
Considering the participants are blindfolded, the experiment demonstrates that touching on the palms is no worse than touching on the fake phone; This implies that the tactile feedback on an imaginary palm interface can achieve a performance similar to the availability of visual clue on the touchscreen of smartphone. The result supports the hypothesis that tactile feedback improves the task performance. After proving that the tactile cue is relevant to task performance, the next important question is about the importance of the tactile sources, that is, active touch and passive touch \cite{[28]Gustafson:2013:UPI:2470654.2466114}.  

The second experiment considers three scenarios  (Figure \ref{fig:Palm}). 1) Palm, 2) Fake Palm, 3) Palm with finger cover. The fake palm is used for evaluating the performance when passive touch is removed, while the finger cover is to discover the effects of active touch on the fingertip. The results show that browsing on the fake palm is significantly slower than on a real palm, while in contrast there is no significant performance gap between touching the real palm with or without a finger cover (tactile sense exists or not). Consequently, the experiment gives evidence that the tactile cue comes from the passive tactile sense (from the palm), instead of the active one (on the fingertip). 

PalmGesture \cite{[45]Wang:2015:PUP:2785830.2785885} is an example of eyes-free interaction using the palm as an interaction surface. It is an implementation based on the findings of an imaginary interface. The interaction highly depends on the tactile cue on the palm of one hand (passive touch), while a finger of another hand acts as the stylus (active touch). The finger performs stroke gestures on the palm, and the user does not require any visual attention on the palm. The proof-of-concept system consists of an infrared camera mounted on the user's wrist, which detects touch events on the palm. The user can enter text by drawing single-stroke Graffiti characters, as well as trigger an email list by drawing an envelope symbol on the palm.
\paragraph{Forearm as surface}
The forearm interface, analogous to the finger-to-palm interaction, can be divided into two approaches. First, widgets or menus are projected onto the surface of the forearm as a visual clue, and the user touches the forearm and obtains a tactile clue. Another approach is eyes-free interaction that solely depends on a tactile clue. The forearm serves as a 'trackpad' and the user rubs across the forearm. 

\begin{figure}
  \includegraphics[width=1\textwidth]{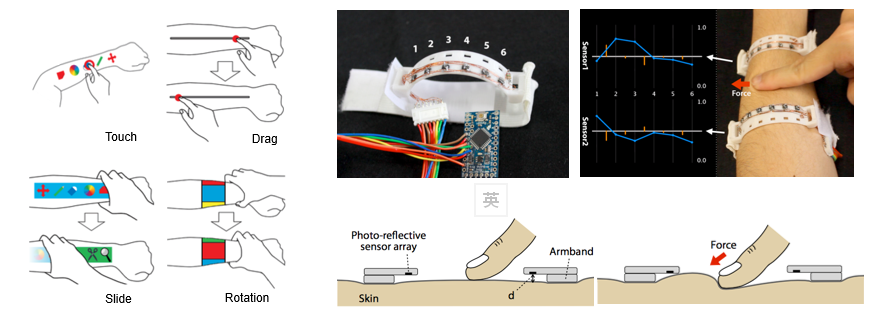}
  \caption{Forearm Widget \cite{[34]Azai:2017:SMM:3027063.3052959} (Left), and hardware configurations of SenSkin \cite{[46]Ogata:2013:SAS:2501988.2502039} (Right)}
  \label{fig:forearm and SenSkin}
\end{figure}

The finger-to-forearm interaction requires either optical or vibration sensors mounted on the arm. As mentioned, Skinput \cite{[33]Harrison:2011:SAS:1978542.1978564} can be applied in finger-to-forearm interaction as long as the projected virtual interface is located on the forearm. Azai et al. \cite{[34]Azai:2017:SMM:3027063.3052959} designs a menu widget on the forearm for smart glasses (Figure \ref{fig:forearm and SenSkin}). Due to the latest development in augmented reality smart glasses such as Microsoft Hololens having a bigger field of view (FOV), the widgets can be fully displayed on the forearm. Four types of interactions are designed for forearm widgets \cite{[34]Azai:2017:SMM:3027063.3052959}, which are Touch, Drag, Slide, and Rotation. The interactions on the forearm are detected by infrared sensors mounted on the top of the head-worn computer. Touch and drag interactions are suitable for item selection and controlling a scrolling bar. Slide means one hand slides from the wrist to the elbow of another hand, and the menu switches accordingly. Rotation is designed for adjusting parameters on the widget such as increasing the volume of a music player. In SenSkin \cite{[46]Ogata:2013:SAS:2501988.2502039}, photo-sensitive sensors can sense any force exerted on the forearm, such as pull, push and pinch on the skin (Figure \ref{fig:forearm and SenSkin}). PUB \cite{[47]Lin:2011:PPU:2047196.2047259} converts the user's forearm into a touch interface by using ultrasonic sensors. SenSkin \cite{[46]Ogata:2013:SAS:2501988.2502039} and PUB \cite{[47]Lin:2011:PPU:2047196.2047259} allow eyes-free interaction and are mainly driven by tactile cues, while Skinput \cite{[33]Harrison:2011:SAS:1978542.1978564} and forearm widget \cite{[34]Azai:2017:SMM:3027063.3052959} offer both visual and tactile cues on the forearm.

\paragraph{Finger as surface}
Finger can be viewed as a part of palm-based interaction. We separate them with the following reasons. We discuss the thumb-to-fingers interaction in this sub-section. It is the subtle movement of the thumb on the index and middle fingers \cite{[37]Huang:2016:DDT:2858036.2858483}; The finger-to-palm interaction has been discussed thoroughly in the previous sub-section.

\begin{figure}
  \includegraphics[width=1\textwidth]{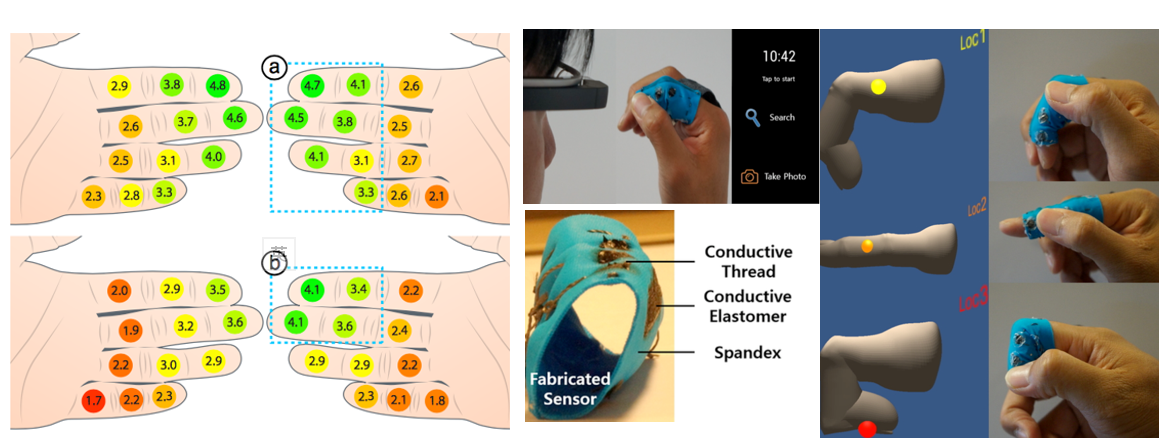}
  \caption{Huang et al's \cite{[37]Huang:2016:DDT:2858036.2858483} suggested zone for thumb-to-finger interaction (Left) and TiMMi \cite{[41]Yoon:2015:TFT:2677199.2680560} (Right)}
  \label{fig:thumb-to-finger}
\end{figure}

The space \cite{[48]Kuo} and coordination \cite{[49]LI2007502} between the thumb and other fingers are crucial to the design of thumb-to-fingers interaction. Huang et al. \cite{[37]Huang:2016:DDT:2858036.2858483} studies the possibility of designing the button (tap gesture) and touch (stroke gesture) widget under the scenario of thumb-to-fingers interaction. The comfortable reach between thumb and other fingers are investigated in their study. The results (Figure \ref{fig:thumb-to-finger}) are as follows. Regarding the button widget, participants prefers to touch on the 1st and 2nd phalanx of the index, middle, and ring fingers, as well as the 1st phalanx of the little finger. As for the touch widget, only the 1st phalanx and 2nd phalanx of the index finger and middle finger are the areas of comfortable reach. Their study also indicates that participants prefers stroke movements because larger movements improve physical comfort. The above findings suggest that the 1st and 2nd phalanx of the index and middle fingers are considered as the ideal area for thumb-to-fingers interaction.
.

\begin{figure}
  \includegraphics[width=1\textwidth]{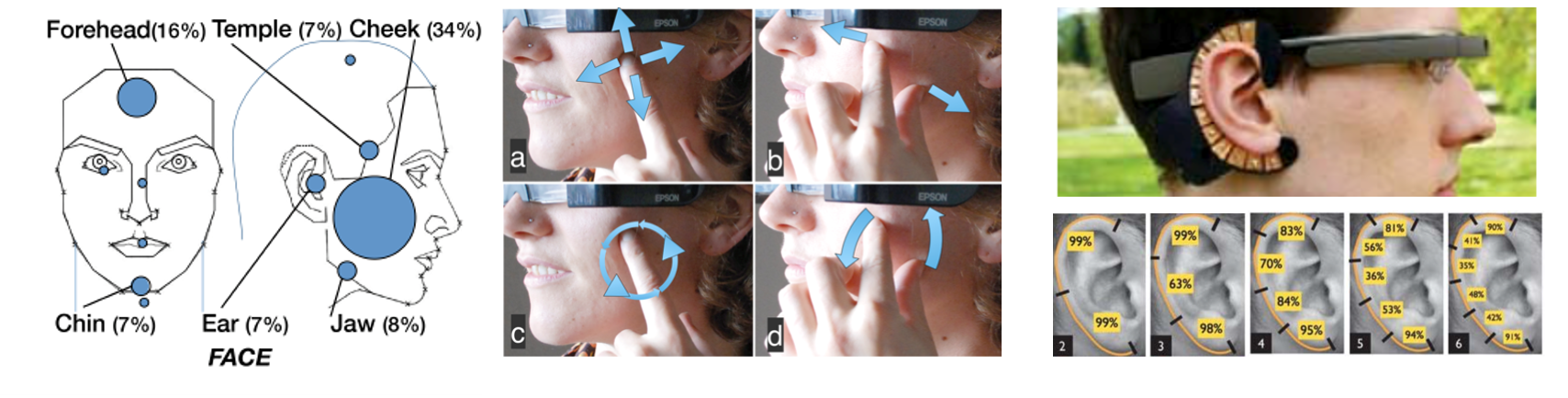}
  \caption{User's preference of facial area \cite{[29]Serrano:2014:EUH:2611247.2556984} (Left), Finger-to-face gestures (Middle) and EarPut \cite{[30]Lissermann:2013:EAB:2468356.2468592} (Right)}
  \label{fig:Face and Ear}
\end{figure}

Furthermore, the implementation of thumb-to-finger interactions are as follows. TiMMi \cite{[41]Yoon:2015:TFT:2677199.2680560} is a flexible surface enclosing the index finger that forms a ring-like device (Figure \ref{fig:thumb-to-finger}). It achieves multi-modal sensing areas between the thumb and index finger. The surface can capture gestures when the thumb exerts forces on the index finger. As the surface is slim and flexible, the press on the surface can give a tactile cue to the user. TiMMi is a rudimentary prototype, while iSkin \cite{[40]Weigel:2015:IFS:2702123.2702391} is a mature prototype ready for commercialization. Likewise, iSkin is a thin, flexible and stretchable overlay on the user's skin. It encloses the index finger, and senses the touch from the thumb. The incredibly thin layer enables the user to receive tactile feedback. Additionally, the remarkable feature of iSkin is that the appearance of iSkin is customizable and aesthetically pleasing, and hence achieves higher social acceptance. According to the indicative examples of iSkin, the layers can be extended to other body surfaces such as forearm, palm, face, and so on. FingerPad \cite{[59]Chan:2013:FPS:2501988.2502016} allows the user's thumb to perform pitch gesture on the 1st phalanx of the index finger, in which magnetic sensors are positioned on the nail of the index finger.
\paragraph{Face as surface}
Serrano et al. \cite{[29]Serrano:2014:EUH:2611247.2556984} proposes a hand-to-face input for interacting with the head-worn display including smart glasses. The face is well suited for natural interaction with the following justifications. First, the facial area is touched frequently, which is 15.7 times per hour in the observational experiment \cite{[35]Mahmoud:2011:IHG:2062850.2062879}. Users feel at ease to do subtle interaction on their faces. The frequent touch on the face means that the gesture could be less intrusive and therefore shows a higher level of social acceptance. Second, the hand-to-face interaction has enough space on the facial area for various gestural interactions including panning, pinch zooming, rotation zooming, and cyclic zooming (Figure \ref{fig:Face and Ear}). An example shown in a user study \cite{[35]Mahmoud:2011:IHG:2062850.2062879}, browsing a webpage requires a lot of panning and zooming. Third, likewise for other on-skin interactions, tactile feedback from the facial area can actually orient the user. When tactile feedback is available, eyes-free interaction is also facilitated \cite{[38]Yi:2012:EUM:2207676.2208678}, and hence minimizes the waiting time for visual feedback \cite{[39]Wagner:2013:BDS:2470654.2466170}. Last, the moment of positioning the user's hand on the facial area can serve as a gesture delimiter that informs the gesture system to record a new gesture and thus avoid unintentional activation.

Regarding the ideal facial area, the lower region of the face is suggested and the facial area in front of the eye and mouth should be avoided because gestural inputs in front of these areas will obstruct the user's view (Figure \ref{fig:Face and Ear}). The area on the cheek is highly preferred by the participants \cite{[29]Serrano:2014:EUH:2611247.2556984} because the cheek imitates the large area of the touchpad on smart glasses, which is regarded as an extension of the touch surface from the body of the smart glasses. However, the task performance is subject to the arm-shoulder fatigue, especially when prolonged use, because the hand-to-face actions require lifting the user's arm. Also, some participants do not accept the hand-to-face interaction because excessive touching could mess up their face makeup or finger skin oil will remain on their face. 
.
\paragraph{Ear as surface}
In this survey, the ear is distinguished from the facial area as the description of face-to-hand input is limited to the cheek. Lissermann et al. \cite{[30]Lissermann:2013:EAB:2468356.2468592} proposes a hardware prototype, namely EarPut (Figure \ref{fig:Face and Ear}), which instruments the ear as an interactive surface for touch-based interaction. The user can touch on the ear and accordingly trigger the arc-shaped capacitive touch sensor at the back of the ear for smart glasses input. Similar to hand-to-face input, the advantages of the hand-to-ear input are four: proprioception, natural tactile feedback, eyes-free interaction, and easy access. 

In comparison with hand-to-face input, the surface area of the ear is relatively small and not flat, e.g. ear helix. The participants prefer to divide the ear into a maximum of four areas. This means the sole reliance on touch or tap is not enough for various interactions. Proposed gestures include touch gestures (such as single tap, slide on ear, and multi-touch on the ear), grasp interactions (e.g. bending the ear, pulling the ear lobe, and covering the entire ear), as well as mid-air gestures. However, the social acceptance of the proposed gestures is not evaluated in their work. The touch gestures on the ear can be considered as an analogous example of hand-to-face input and thus it is suitable for use in a public area. However, the acceptance of grasp interactions, blending the ear especially, and mid-air gestures next to the ear are still questionable.

\subsection{Touchless inputs}
Regarding the touchless inputs, smart glasses users make gestural input mid-air and receive visual clues from the optical display on the smart glasses. The touchless input can be classified into two categories: Hands-free and Freehand interactions. Hands-free interaction can be made by the movements of the head, gaze, voice and tongue, while freehand interaction focuses on mid-air hand movements for gestural input. 

\subsubsection{Hands-free interaction}

Hands-free input is one of the most popular categories in the domain of interaction techniques. It enables users to perform hands-free operations on smart glasses. That is, interaction between users and smart glasses involves no hand control. In the wide body of literature, hands-free interaction techniques include voice recognition, head gestures, and eye tracking. In addition, tongue gestures have been studied in recent years.
\paragraph{Voice recognition}
This technology has been deployed in smart glasses and becomes the major input method for Google Glass and Microsoft Hololens. However, it might be inappropriate in shared or noisy environments, for example, causing disturbance and obtrusion \cite{[69]7524542}, disadvantages to mute individuals, accidentally activated by environmental noise, and less preferable than the input approaches by body gestures and handheld devices \cite{[70]Kollee:2014:EGI:2659766.2659781}. 
\paragraph{Head movement}
Head-tilt gestures are mainly driven by built-in accelerometers and gyroscopes in smart glasses. This technology is applicable to text input \cite{[71]Jones:2010:GAG:1753326.1753655}, user authentication \cite{[69]7524542} as well as game controller  \cite{[89]Wahl:2015:USE:2800835.2800914}. Glass Gesture \cite{[69]7524542} utilizes both accelerometer and gyroscope in smart glasses to achieve high input accuracy, in which a sequence of head movements is regarded as authentication input. In \cite{[89]Wahl:2015:USE:2800835.2800914}, users can control the movement (up, right, left, down) of the characters in a Pac-Man game by head movement. However, head movements cannot be considered as the major input source due to the ergonomic restriction of users moving their heads for long-periods of gaming. 
\paragraph{Gaze movement}
Gaze movement can instruct the cursor movement for pointing tasks \cite{[93]Ware:1986:EET:29933.275627}, for instance, choosing an object with an eye gaze \cite{[19]Slambekova:2012:GGB:2407336.2407380,[94]Toyama:2014:MRH:2557500.2557528,[114]Bace:2016:UUA:2999508.2999530}, text input based on Dasher writing system \cite{[90]Tuisku:2008:DDA:1344471.1344476}, and recognizing objects with eye gaze in augmented reality \cite{[22]Toyama:2012:GGO:2168556.2168570}. Gaze interactions have been proposed for head-mounted displays \cite{[21]Shimizu:2016:EMI:2968219.2968274,[72]Schuchert:2012:SVA:2401836.2401852} and smart glasses \cite{[89]Wahl:2015:USE:2800835.2800914}. Slambekova et al. \cite{[19]Slambekova:2012:GGB:2407336.2407380} have designed multi-modal system for fast object manipulation of virtual contents. Gaze input acts as a mouse cursor that chooses objects and simultaneously hand gestures performs object manipulation such as translation, rotation and scaling. The system well utilizes the characteristics of eye and hand. Gaze interaction can catch the target object quickly and human hands have a high degree of freedom (DOF) that enables manipulating objects in diverse manners. Toyama et al. \cite{[94]Toyama:2014:MRH:2557500.2557528} utilizes gaze movement to select the targeted text for translation on the optical display of smart glasses. In UbiGaze \cite{[114]Bace:2016:UUA:2999508.2999530}, users can embed visible messages into any real-world object and retrieve such messages from those objects with the assistance of gaze direction which indicate where the users are looking in the surrounding physical environment.

Eye movement is a natural and fast input channel, in which only slight muscle movement is involved, but it has major drawbacks, for instance, they are error-prone and suffer from excessive calibration, and the eye-tracking hardware is not available in smart glasses \cite{[91]Bulling:2012:GIP:2212776.2212428}. The performance of gaze input can be further improved by considering haptic feedback. Kangas et al. \cite{[92]Kangas:2014:GGH:2556288.2557040} studies the effect of vibro-tactile feedback from a mobile device as a confirmation of gaze interaction. The results show that the task completion time is shortened when the vibro-tactile feedback is available, and the participants feel comfortable due to reduced uncertainty. Nonetheless, the eye-tracking technology for smart glasses will not be popular for the next several years because the price of the tracker is no less than a few hundred dollars.
\paragraph{Tongue movement}
The tongue machine interface is usually proposed for paralyzing injuries or medical conditions which retain the use of their cranial nerves \cite{[31]Saponas:2009:OST:1622176.1622209}. The locations of sensors can be either intrusive \cite{[31]Saponas:2009:OST:1622176.1622209,[36]Zhang:2014:NTM:2556288.2556981} or non-intrusive \cite{[73]Goel:2015:TUW:2702123.2702591}. Saponas et al. \cite{[31]Saponas:2009:OST:1622176.1622209} places infrared optical sensors inside the user's mouth to detect the tongue movement. Four simple gestures (back, front, left, right) are achieved with 90\% accuracy. Zhang et al. \cite{[36]Zhang:2014:NTM:2556288.2556981} locates electromyography sensors on the user's chin to detect the muscle changes driven by tongue gestures. Two additional gestures (protrude and rest) are designed in [36] with 94.17\% accuracy. Tongue-in-Cheek \cite{[73]Goel:2015:TUW:2702123.2702591} has a system that uses 10 GHz wireless signals to detect different facial gestures in four directions (up, right, left, down) and two modes (tap and hold). It detects the facial movement on cheeks driven by moving different parts of the mouth: touching of tongue against the inside of the cheeks, puffing the cheeks, and moving the jaws. The total 8 gesture combinations achieve 94.30\% accuracy. Even though the tongue interface can achieve highly accurate detection, it lacks considerations for a complicated interface. Only simple gestures are demonstrated in the testing scenarios like Tetris \cite{[31]Saponas:2009:OST:1622176.1622209}. It is very likely that the current works on tongue machine interface are not ready for interactions in augmented reality.

\subsubsection{Freehand interaction}

Although various hands-free techniques are proposed in the literature, there is no evidence showing that hands-free input with smart glasses outperforms other interaction techniques such as freehand interaction. As reported by Zheng et al. \cite{[74]Zheng:2015:ETM:2702123.2702305}, human beings are good at adapting to various conditions whether or not their hands are occupied by instruments or tasks or not. In other words, performing hands-free operations may not be the necessary condition in the design of interaction techniques and thus freehand interaction involving hand gestures is not inferior to hands-free input. A usability study \cite{[24]Tung:2015:UGI:2702123.2702214} also found that the gestural input is preferable to on-body gestures and handheld devices especially in an interactive environment. 

\begin{figure}
  \includegraphics[width=0.5\textwidth]{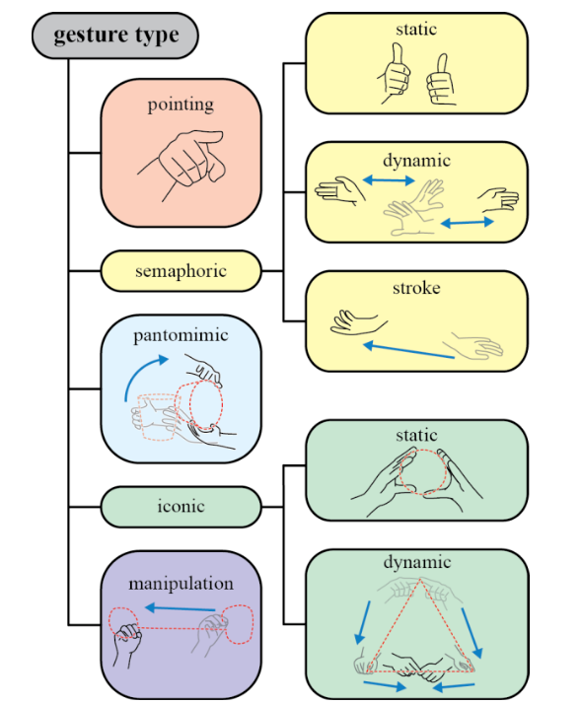}
  \caption{8 types of hand gestures \cite{[97]microsoftresearch}}
  \label{fig:HandGesture}
\end{figure}

Freehand interaction refers to the human-smart glasses interaction driven by hand gestures. Hand gestures can be classified into 8 types \cite{[97]microsoftresearch}: Pointing, Semaphoric-Static, Semaphoric-Dynamic, Semaphoric-Stroke, Pantomimic, Iconic-Static, Iconic-Dynamic, and Manipulation, as shown in Figure \ref{fig:HandGesture}. The followings is a brief explanation of the listed hand gesture types. 
\begin{enumerate}
\item Pointing: Used to select an object or to specify a direction. Pointing can be represented by index finger, multiple fingers, or a flat palm.
\item Semaphoric-Static: Derived meaning from social symbols such as thumbs-up as 'Like' and forward-facing flat palm as 'Stop'. The symbols can be carried out with one or both hands and be directed to the camera without movement.
\item Semaphoric-Dynamic: Added temporal aspect on the Semaphoric-static. Clock-wise rotation motion means 'Time is running out'. 
\item Semaphoric-Stroke: Similar to Semaphoric-dynamic, but an additional constraint of a single dedicated stroke is considered. Examples can be 'Next/Previous Page'.
\item Pantomimic: Considered a single action of mime actor to illustrate a task, for example, grabbing an object, as well as moving and dropping an object.
\item Iconic-Static: Pertaining to an icon, for instance, making an oval by cupping two hands together.
\item Iconic-Dynamic: Added temporal aspect on Iconic-Static. An example is constantly circular hand movement (i.e. drawing a circle).
\item Manipulation: the above gesture types requires a pre-defined time interval to recognize the hand gesture. This type refers to executing a task once the user performs a particular gesture. Considering moving an virtual 3D object, no delay should exist once the mid-air touch on the virtual object is executed and the update of an object's location should be instantly performed in a continuous manner. 
\end{enumerate}
Sensors are necessary for capturing the dynamic movements and static postures of a user's hand. Glove and camera are commonly used for freehand interaction with smart glasses. In this sub-section, we focus on the recent works on smart glasses. 
\paragraph{Glove}
The device is commonly comprised of sensors and inertial measurement units to detect hand gestures and postures. A comprehensive review of the history and advancement of glove-based systems can be discovered in \cite{[55]4539650}. In general, glove-based interaction is applied to hand gestures of the pointing kind \cite{[83]Haque:2015:MPC:2702123.2702133,[96]Hsieh2014} and text input \cite{[95]Rosenberg:1999:CGG:2220393.2220512}. Vulture \cite{[79]Markussen:2014:VMW:2611105.2556964} is a mid-air hand gesture interaction technique for text entry, where the instrumental gloves tracks hand and fingers positions. Myopoint \cite{[83]Haque:2015:MPC:2702123.2702133} contains electromyography and inertial motion sensors detecting arm muscle movement and achieves pointing and clicking through muscle contraction and relaxation. Recent works related to smart glasses interaction are mainly designed for particular considerations such as enhancing social acceptance of gestural inputs \cite{[81]Hsieh:2016:DWH:2858036.2858436}, ubiquitous gaming in mixed reality \cite{[82]Martins:2008:GWI:1409240.1409290}, designed rested posture for long-term use \cite{[101]Guinness:2015:MRT:2788940.2788948}, and supporting tangible augmented reality on physical objects \cite{[107]Simon:2012:MTB:2512125.2512130}.
\paragraph{Camera}
Multiple types, such as RGB, depth, infrared, thermal camera and so on, of Camera enable vision-based approaches are for freehand interaction. The recent works have applied RGB camera \cite{[50]Chan:2015:CEW:2807442.2807450,[76]Huang:2015:UTS:2733373.2806266,[77]4373785,[84]Bailly:2012:SNP:2207676.2208576,[103]10.1109/MOBIQ.2004.1331713}, and depth camera \cite{[56]Rekimoto:2001:GGU:580581.856565,[80]Guimbretiere:2012:BMM:2207676.2208521,[104]6948431}, which are image processing, tracking, and gesture recognition. The components of tracking and recognition can be achieved by mainly two approaches: model-based or appearance-based. Forearm, hand and finger are the target object in the gesture recognition \cite{[102]Moeslund03abrief}. A systematic literature review showing the development of mid-air hand gestures refers to \cite{[99]Groenewald:2016:UMH:3056355.3056398}.

Regarding vision-based freehand interaction with smart glasses, there have been a number of gestural interfaces with diverse purposes. From the early works, we can see hand gestures are applied as a mouse cursor that enables interactions with a 2D interface in the optical display \cite{[80]Guimbretiere:2012:BMM:2207676.2208521,[103]10.1109/MOBIQ.2004.1331713}. As augmented reality owns the prominent features of the integration of virtual contents with the physical environment, hand gesture shows its intuitiveness and convenience in the environment \cite{[106]citeulike:8213624}. Huang et al. \cite{[76]Huang:2015:UTS:2733373.2806266} propose a hand gesture system that facilitates interaction with 2D contents overlaid on physical objects in an office environment. In addition, Heun et al. \cite{[105]Heun:2013:SOU:2468356.2479579} enhances the capability of simple physical objects, such as knobs and buttons, by augmenting a 2D tangible interface on a tangible surface or on top of a physical object. On the other hand, hand gesture systems are designed for manipulating virtual 3D objects in augmented reality. An early work utilizes human hand to substitute for a fiducial marker \cite{[77]4373785}, and another recent work enables barehanded manipulation of virtual 3D objects in augmented reality \cite{[104]6948431}. 

Ubii \cite{[76]Huang:2015:UTS:2733373.2806266} is a gestural interface in which users can perform in-situ interactions with physical objects from a distance, including computers, projects screens, printers, and architecture partitions in an office environment. With the assistance of fiducial markers, users can simply apply hand gestures to complete tasks such as document copying, printing, sharing, and projection display. Kolsh et al. \cite{[103]10.1109/MOBIQ.2004.1331713} proposes an ego-centric view interface that enables users to perform pointing gestures in a 2D interface. In addition, some Iconic-Static gestures are included in their work, for instance, if two open hands are settled for five seconds, the head-mounted camera takes a snapshot. Similarly, Francois and Chan \cite{[80]Guimbretiere:2012:BMM:2207676.2208521} have proposed a multi-finger pinching system that simulates multi-button mouse interaction under depth camera, for instance, pinching gestures with index finger or middle finger invokes left and right clicks, respectively. A marker-less camera tracking system for 3D interface, namely Handy AR \cite{[77]4373785}, uses the hand pose model to substitute the fiducial marker for 3D objects tracking and manipulation in augmented reality. By transforming the palm and fingers on the outstretched hand into the hand pose model, users can manipulate the 3D object by hand rotation and movement in augmented reality. In WeARHand \cite{[104]6948431}, users can select and manipulate virtual 3D objects with their own bare hands in a wearable AR environment, for instance, moving the virtual 3D object from one location to another.

The above works commonly uses a head-worn camera or a camera embedded in smart glasses. Cameras can also be positioned on arms \cite{[56]Rekimoto:2001:GGU:580581.856565,[51]PinchWatch}, fingers \cite{[50]Chan:2015:CEW:2807442.2807450}, shoes \cite{[84]Bailly:2012:SNP:2207676.2208576}, chests and belts \cite{[78]Gustafson:2010:IIS:1866029.1866033}. These approaches using wearable cameras aim to provide subtle interactions for higher social acceptance \cite{[50]Chan:2015:CEW:2807442.2807450,[56]Rekimoto:2001:GGU:580581.856565} and free body movement \cite{[84]Bailly:2012:SNP:2207676.2208576,[78]Gustafson:2010:IIS:1866029.1866033} that prevents gorilla arm \cite{[100]Hincapie-Ramos:2014:CEW:2598784.2602795}. Pinchwatch \cite{[51]PinchWatch} has a wrist-worn depth-camera to capture the thumb-to-palm and thumb-to-fingers interactions. CyclopsRing \cite{[50]Chan:2015:CEW:2807442.2807450} detects the webbing of fingers by a fisheye RGB-camera in which the segmentation of skin color on fingers can produce a 2D silhouette for gesture recognition. Shoe-Sense \cite{[84]Bailly:2012:SNP:2207676.2208576} has an upward-oriented optical sensor installed on a shoe. Users can make various two-armed poses in triangular form and the sensor can read the triangular arm gestures. Gustafson et al. \cite{[78]Gustafson:2010:IIS:1866029.1866033} proposes an imaginary mid-air interface for wearables without touchscreens. The camera on the user's chest owns a wide perspective that captures the user's hand movement and accordingly allows input such as graffiti characters, symbol and curves.

While hand gestural interaction has compelling features such as natural and intuitive interaction, mouse and touch interaction outperforms the hand gestural interaction for fast repetitive tasks. An exploratory study \cite{[85]Sambrooks:2013:CGT:2541016.2541066} shows the comparison between gestural, touch, and mouse interaction in the WIMP paradigm with Fitt's Law \cite{[68]MacKenzie:1992:FLR:1461854.1461857}. The results indicate that gestural interaction suffers from inaccurate recognition (hit-to-miss ratio is 1:3), poor performance time due to potential unfamiliarity with hand gesture library, and muscle fatigue. Another study also aligns with these findings \cite{[86]Pino:2013:UKP:2530824.2530864}. Additionally, gestural interaction requires relatively long dwelling time compared with mouse or touch interaction, and consequently an intensive task is not appropriate. The user needs to hold the posture for a period of time and this problem is regarded as the Midas problem\cite{[120]Istance:2008:SCM:1344471.1344523}, in which guessing the gesture initiation and termination are consuming and erroneous\cite{[119]Chen:2016:PMM:2851581.2892492}. It is concluded that gestural interaction is slower and harder to use than direct pointing interaction in a 2D interface.

A midpoint on the spectrum between Direct pointing and Semaphoric gesture should be taken into consideration. Some gesture types for 2D interfaces, such as Pantomimic and Iconic gestures, are less than ideal as discussed. Therefore, gesture type towards barehanded direct pointing \cite{[75]Ren:2013:FGT:2465958.2465966} is a potentially fruitful direction for 2D interface interaction on smart glasses. Moreover, direct pointing or manipulation are analogue to the touch interface on a smartphone, that is, touchscreen, but the mouse interaction is not available on smart glasses and the visual content is no longer touchable. Therefore, pointing gestures become a viable option for 2D interfaces, for instance, Heo et al. \cite{[88]7066537} proposes a vision-based pointing gesture system by detecting the number of fingertip, instead of identifying the silhouette of the hand posture. More importantly, virtual 3D objects are always involved in augmented reality. Gesture types such as Semaphoric-stroke should be considered because of its natural and intuitive interaction \cite{[87]REN2013101}.

\section{Existing research efforts and trend}
\begin{figure}
  \includegraphics[width=1\textwidth]{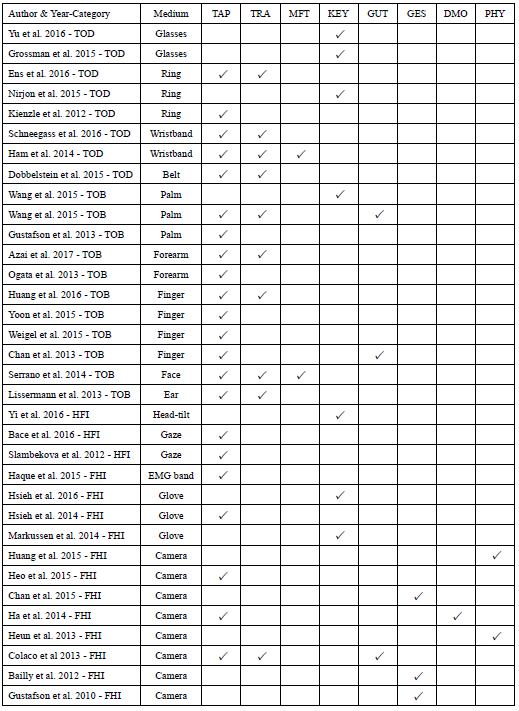}
  \caption{The coverage of research efforts on smart glasses interaction}
  \label{fig:coverage_table}
\end{figure}

In this section, we evaluate the interaction approaches from the perspective of interaction goals, including spectacle frame of smart glasses, rings, wristbands, belts, body surfaces, body movements, gloves and cameras. Based on the proposed classification system for touch and touchless input, their input abilities are discussed. According to the characteristics and features of the identified works in the previous section, we choose and compare more than 30 research works relevant to smart glasses interactions in recent years. All these works representing their categories \textbf{\textit{(TOD}}: Touch-on-device,\textbf{\textit{ TOB}}: Touch-on-body,\textbf{\textit{ HFI}}: Hands-free interaction, and \textbf{\textit{FHI}}: Freehand interaction) are designed for various interaction goals including manipulating an item and a scrolling bar inside a 2D interface, selecting a key on a virtual keyboard, writing graffiti words and unistrokes in text entry systems, manipulating 3D objects, interacting with a physical object in augmented reality, as shown in Figure \ref{fig:coverage_table}. In the table, the interaction goals are summarized into 8 types, as the followings: \textbf{\textit{TAP}}: single-tap gestures for operating items (e.g. select and drag a button or a menu), including single-tap and tap-and-hold, \textbf{\textit{TRA}}: single-finger gestures that produces a trajectory for stroke inputs (e.g. swipe for switch between pages as well as scroll up/down, drawing a circle or envelope), \textbf{\textit{MFT}}: multi-finger touch gestures such as zooming in/out, cyclic gestures, \textbf{\textit{KEY}}: selecting keys on a virtual keyboard and other non-stylus based text entry techniques, \textbf{\textit{GUT}}: stylus based (e.g. graffiti or unistroke) inputs for text entry systems, \textbf{\textit{GES}}: hand gestural commands, a total of eight types as discussed in Section 3.2.2, \textbf{\textit{DMO}}: direct manipulations on virtual three-dimensional objects (e.g. rotation, translation), \textbf{\textit{PHY}}: interacting with a physical environment in augmented reality.

From the results shown in Figure \ref{fig:coverage_table}, it is easy to recognize the touch input is mainly designed for tap and swipe gestures (\textbf{\textit{TAP}} and \textbf{\textit{TRA}}), as well as text entry system (\textbf{\textit{KEY}} and \textbf{\textit{GUT}}). The tap and swipe gestures are what commonly used for smartphone interface supporting multitudinous tasks. The on-device and on-body touch inputs aim to provide alternatives input approaches for smart glasses. Except those research works solely focusing on text inputs (Grossman et al. 2015, Yu et al. 2016, Nirjon et al. 2015, Wang et al. 2015), single-tap and stroke-based gestures are commonly available on touch inputs. When a larger surface is available on an external device or a body skin surface, multi-finger touch gestures are proposed accordingly (\textbf{\textit{MFT}}). For example, the wristband proposed by Ham et al. \cite{[57]Ham2014} has a phone-sized touch interface, and the pioneer work of finger-to-face interaction proposed by Serrano et al. \cite{[29]Serrano:2014:EUH:2611247.2556984} utilizes the considerably large surface on cheeks. Additionally, we observe that touch interfaces are responsive and accurate designs that sufficiently supports various tasks. Consequently, the touch input doesn't necessarily support all the eight types of gestural inputs (\textbf{\textit{GES}}). Interestingly, the gesture types such as Semaphoric-Stroke and Iconic-Static are also convenient on the 2D sensible touch surface. Instead of performing mid-air hand swipe, the user can draw a stroke on a touch-sensitive interface \cite{[52]Schneegass:2016:GUT:2971763.2971797}. Similarly, the user can draw an icon of envelope to trigger an email application \cite{[45]Wang:2015:PUP:2785830.2785885}. 

Four noticeable trends are identified in the existing works of touch input, as follows. First, research studies of on-body interaction focus on upper limbs \cite{[43]Weigel:2014:MTU:2556288.2557239} and facial areas \cite{[29]Serrano:2014:EUH:2611247.2556984}. On-body interaction requires sensor channels detecting either infrared light \cite{[45]Wang:2015:PUP:2785830.2785885} or vibration exerted on skin surfaces \cite{[47]Lin:2011:PPU:2047196.2047259}. The existence of additional sensor arrays in external device merely serves as a detector of on-skin interaction. Inertial measurement units can possibly integrate with the external device for more variations of gestural inputs. Second, the research efforts have considered various forms and sizes of external devices and a wide range of sensing capabilities. Considering the finger-worn device as example, the device can be made in the form of traditional ring, distal addendum, whole finger addendum, fingernail addendum, finger sleeve, thumb addendum \cite{[58]Shilkrot:2015:DDC:2830539.2828993}. In addition, the capabilities of sensors can influence the gesture library on the ring, for instance, detecting the degree of the bending of the finger muscle can enrich the possible number of gestures in the gesture library on the limited touch-sensitive surface \cite{[62]Ogata:2012:IIR:2380116.2380135}. Third, there are no evident restriction on the recommended size of the touch surface. The selection of surface size is commonly justified by the functionality and the social acceptance. The common norms are that a larger surface, such as wristbands, finger-to-forearm, finger-to-face and finger-to-palm interactions, can support more comprehensive gestural inputs. For example, SenSkin \cite{[46]Ogata:2013:SAS:2501988.2502039} supports single tap (\textbf{\textit{TAP}}), drawing trajectory (\textbf{\textit{TRA}}), and multi-finger gesture (\textbf{\textit{MFT}}). In contrast, a smaller surface such as finger-worn devices and thumb-to-finger interaction usually supports simpler gestural inputs (\textbf{\textit{TAP}} and \textbf{\textit{TRA}}). In addition, the smaller surface supports subtle and inattentive one-hand interactions, which is more favorable in terms of social acceptance. The larger surface requires the two-hand interaction and the discernible body movement, which can raise unfavorable attention from the surroundings. From some recent works, we observe that the smaller surface attempts to expand its functionality. For instance, text entry is commonly proposed on a larger surface like finger-to-palm interaction \cite{[42]Wang:2015:PUP:2785830.2785886}. A finger-worn device has demonstrated the interaction potentials beyond the tap and swipe gestures. In TypingRing \cite{[121]Nirjon:2015:TWR:2742647.2742665}, the finger-worn device achieves a typing speed of 6.26 - 10.44 word per minutes, while finger-to-palm interaction such as PalmType \cite{[42]Wang:2015:PUP:2785830.2785886} can only achieve 4.7 word per minute. Another example can be the text entry system using the graffiti word. Both the finger-to-thumb \cite{[59]Chan:2013:FPS:2501988.2502016} and Palm Gesture \cite{[45]Wang:2015:PUP:2785830.2785885} allow user to write graffiti words on skin surfaces. These works show that the surface size is not necessarily a trade-off with the comprehensiveness of design functions. Last, the eyes-free interaction is an important feature of the touch interaction, which is supported by the existence of tactile cue. The benefits are discussed in previous section. 

On the other hand, touchless input demonstrates distinguishable input characteristics from touch input. Hands-free interaction such as head and gaze movement provides very limited functions, that is, they are designed for micro-interaction such as a short duration of authentication inputs \cite{[69]7524542} and locating a few items in augmented reality \cite{[114]Bace:2016:UUA:2999508.2999530}. The freehand interaction enabled by gloves and electromyography (EMG) wristbands can achieve item selections (\textbf{\textit{TAP}}) on a virtual interface, and mid-air text input through sensing the finger movement (\textbf{\textit{KEY}}). We consider that vision-based freehand interactions are interested primarily in the manipulation of a 3D objects (\textbf{\textit{TAP}} and \textbf{\textit{DMO}}) and a physical environment in augmented reality (\textbf{\textit{PHY}}). These works emphasize the unique characteristics of hand gesture, i.e. intuitiveness and naturalism of direct manipulation on a virtual 3D objects and a physical environment in augmented reality. They also support single tap and drawing trajectory in mid-air that enable manipulating a virtual icons and switching pages in a virtual 2D interface, regardless of the fact that touch input has better performance in terms of accuracy, speed and repetitiveness \cite{[85]Sambrooks:2013:CGT:2541016.2541066}. Interestingly, we discover that the research efforts of vision-based approaches for text input systems are commonly regarded as sign languages \cite{[99]Groenewald:2016:UMH:3056355.3056398}, which is deliberately designed for people with special needs. However, the iconic-static sign language is not appropriate for the purpose of intensive text entry because it suffers from long dwelling time of recognizing every single hand sign \cite{[120]Istance:2008:SCM:1344471.1344523} and hence unproductive input speed. In addition, the mid-air tap on the virtual keyboard often appears in the usage examples of these works but they also suffer from the accumulated dwelling time of recognizing tap gestures on the keys of a virtual keyboard. From the above existing research works, we find that the vision-based freehand interactions are distinctive assets on direct manipulating virtual 3D objects and physical environment, however, the concerns on text entry makes vision-based freehand interactions cannot be an all-rounded approach.

Multi-modal input is one of the prominent trends in the existing work of freehand interaction for smart glasses. First, there exist several pioneer research works combining the benefits of touch and touchless input. Ens et al. have applied finger-worn device to reinforce the subtle movement on small items in a 2D interface, as freehand interaction on small items is lack of precision and fatigue-prone \cite{[25]Ens:2016:CRI:2983310.2985757}. In the system design, hand gestures are assigned to locate large items, such as windows and menus, while the finger-to-ring interaction is responsible for subtle operations on the large items, like relocating a window and changing some parameters in a scroll bar. Instead of having external touch interface on ring devices, Zhang et al. \cite{[118]Bai:2014:UHG:2559206.2581371} utilize the touch interface on the spectacle frame of smart glasses. Second, multi-input modal has been considered to alleviate the issue of dwelling time. Yu et al. \cite{[119]Chen:2016:PMM:2851581.2892492} have exploited the use of electromyography (EMG) sensor on commercial smart wristband to minimize the idle time of detecting the initiation and termination of intended gestures. Another trend is designing low-power hand gesture systems. MIME \cite{[26]Colaco:2013:MCL:2501988.2502042} applies the hybrid processing of image information captured from both RGB and depth cameras. Optimized arrangement of the image sources can achieve both accurate and low-power gesture detection. The depth channel operates intermittently to enhance the performance of color-based detection of hand gesture and avoids intensive uses of power-consuming depth channel.

In conclusion, touch input and freehand interaction are the most popular research topics in smart glasses interaction. Figure \ref{fig:coverage_table} shows the coverage of interaction goals by the proposed four categories, in which touch input (\textbf{\textit{TOD}} and \textbf{\textit{TOB}}) shows promising interaction capabilities in 2D interfaces and vision-based freehand interaction (\textbf{\textit{FHI}}) demonstrates intuitive and natural interactions with virtual 3D objects and physical environment in augmented reality. We envision the trend of combining the touch and touchless inputs have great potentials to smart glass interaction and meanwhile the boundary between freehand interaction and touch input will become ambiguous.

\section{Interaction challenges on smart glasses}
So far we have discussed four categories of interaction approaches that are important to smart glasses interaction. It is essential to note that these categories are research areas that need to be explored further and significantly. We have also provided a coverage of research efforts that readers can use to investigate and fill the performance gap among the interaction approaches. In this section, we highlight a number of challenging problems in smart glasses interaction. The reader may consider the below challenges as some design directions and guidelines for devising new interaction approaches on smart glasses.
\paragraph{Hybrid user interface on smart glasses}
Smart glasses are mobile device and its goal is to deliver an interface of augmented reality to users. Augmented reality involves superimposing interactive computer graphics images onto physical objects in the real world \cite{[110]Poupyrev:2002:DGA:619073.621931}. The virtual contents on the optical display can be represented by the taxonomy of 2D and 3D objects. This combination of virtual 2D and 3D contents can be regarded as hybrid user interface \cite{[106]citeulike:8213624}. The interactions with the virtual contents in the hybrid user interface creates a more intricate and complex scenario than what we have seen on smartphones. The virtual 2D contents refers to the operations on icon, menus and windows in 2D interfaces, for instance, selecting an object \cite{[60]Kienzle:2014:LAI:2642918.2647376}, drawing a trajectory \cite{[25]Ens:2016:CRI:2983310.2985757}, and illustrating a symbolic icon on two-dimensional space \cite{[52]Schneegass:2016:GUT:2971763.2971797}. The 3D contents refer to direct manipulation on virtual 3D objects and augmented information superimposed on physical objects, such as translation and rotation of 3D objects \cite{[104]6948431}  and instructing a printer for printing jobs \cite{[76]Huang:2015:UTS:2733373.2806266}. In the works we surveyed, the virtual 2D and 3D contents can be matched into the eight types of interaction goals mentioned in Section 4. In general, virtual 2D contents can be effectively managed by the types of \textbf{\textit{TAP}}, \textbf{\textit{TRA}}, \textbf{\textit{MFT}}, \textbf{\textit{KEY}} and \textbf{\textit{GUT}}, while virtual 3D contents can be handled by the remaining types of \textbf{\textit{GES}}, \textbf{\textit{DMO}} and \textbf{\textit{PHY}}. Figure \ref{fig:coverage_table} has clearly shown that no existing works can provide a full coverage of eight interaction goals. This obvious gap depicts an immense opportunity for researchers to develop comprehensive approaches for interaction in hybrid user interface on smart glasses. 
\paragraph{Towards higher coverage of interaction goals}
From the results in Figure \ref{fig:coverage_table}, touch input mainly aids the interactions with virtual 2D contents (\textbf{\textit{TAP}}, \textbf{\textit{TRA}} and \textbf{\textit{MFT}}) and text entry (\textbf{\textit{KEY}} and \textbf{\textit{GUT}}). Touchless input dominates the interaction with virtual 3D contents (\textbf{\textit{GES}}, \textbf{\textit{DMO}} and \textbf{\textit{PHY}}). The reasons are as follows. First, the interaction with 2D interface usually needs fast repetition and accurate input, in which high dexterity of fingers on finger-to-device/body interfaces poses more advantageous than the movement of larger body part in mid-air, and the mid-air movement of larger body part (e.g. head and head gestures) is criticized by the lack of precision and prone to fatigue \cite{[25]Ens:2016:CRI:2983310.2985757,[100]Hincapie-Ramos:2014:CEW:2598784.2602795}. As a result, touch input has demonstrated higher input performance than touchless input in terms of interactions of 2D interface and text entry systems \cite{[68]MacKenzie:1992:FLR:1461854.1461857,[86]Pino:2013:UKP:2530824.2530864}. On the other hand, freehand interactions have exhibited its capability in 3D interface among the touchless inputs. Most of the users prefers interaction of 3D objects with hand gesture more than touch-based approaches because users agreed that performing gesture in front of face is natural and straightforward \cite{[24]Tung:2015:UGI:2702123.2702214}. The results can also be justified by the intuitiveness of hand gesture. Hand gesture enables users to direct manipulate the virtual 3D contents, for instance, rotation and translation can be done by simply rotating the wrist and swiping the hand, respectively. In comparison, touch input is less straightforward. For instance, the user first rubs on a touch surface to locate the targeted 3D object, and afterwards draws a circle on the touch surface to rotate the targeted 3D object. In order to achieve a higher coverage of interaction goals in hybrid user interface, it is worthwhile to judiciously consider exploiting both the touch-based and touchless gestures. 
\paragraph{Building all-rounded interaction approaches}
In order to devise interaction approaches on smart glasses fulfilling the aforementioned interaction goals, one possible solution is to make the touch and touchless inputs to tackle its interaction challenges. Touch input can provide more intuitive gestures for the interaction with virtual 3D contents, while touchless input has to fill its gap in tasks requiring fast repetition. Another possible solution is to mingle the touch and touchless inputs together. We envision this assortment of input methods is a like-wise interaction as the multi-modal input appearing in touchscreen computer, e.g. Microsoft Surface. As discussed, exploiting the combination of touch and touch inputs can gain benefits of both inputs, as follows. Hand gesture is ideal for fast, coarse and convenient manipulation of virtual 3D objects, while the operations on virtual 2D interfaces can be fulfilled by touch surfaces that are suitable for precise and longer usage, such as surfing on web browser, selecting items in a widget menu, as well as inputing texts. 
 
According to the surveyed works, we anticipate that the augmented reality on smart glasses would consist of a number of virtual large contents including menus, widgets, windows and 3D objects \cite{[104]6948431}. Inside the large contents, there exist some small contents such as buttons, icons and scroll bars in menus/widgets, and adjusting parameters of 3D objects \cite{[76]Huang:2015:UTS:2733373.2806266}. Under this circumstance, users could first locate the large contents by fast and coarse hand gestures, and subsequently manipulate the small contents with subtle and repetitive touch inputs \cite{[25]Ens:2016:CRI:2983310.2985757}. We here elicit possible configurations for building comprehensive interaction approaches. The touch interface can be designed as a companion device to work complementary with touchless input. Here are two illustrative configurations. 1) touch interface on finger-worn device and vision-based freehand interaction, and 2) haptic glove equipped with touch-sensible textile for touch input, and embedded sensor (in the glove) supporting freehand interaction. Building multi-modal inputs using companion devices may circumvent the obstacles of interaction with smart glasses. To conclude, we see the strengths and weaknesses of input approaches. A variety of interaction potentials can be achieved by considering various combinations of input approaches. These combinations aim at supporting natural and fast interaction for augmented reality on smart glasses. The multi-modal inputs on smart glasses would be one of the most exciting research areas for further investigation. In the rest of this section, several key design factors for the multi-modal inputs on smart glasses are highlighted.
\paragraph{Form size for wide-ranging coverage}
When multi-modal inputs are considered, the choice of inputs can influence the comprehensiveness for the coverage of interaction goals \cite{[60]Kienzle:2014:LAI:2642918.2647376,[53]Dobbelstein:2015:BUT:2702123.2702450,[52]Schneegass:2016:GUT:2971763.2971797,[39]Wagner:2013:BDS:2470654.2466170}. For example, the coverage of interaction goals can be influenced by the size of touch-sensitive area on touch input device. The skin surfaces on forearms and palms as well as wristbands are considered as large interaction areas, which can be regarded as a full-sized trackpad for various missions (e.g. drawing trajectory and text entry). In comparison, the thumb-to-finger interaction and finger-worn devices have very limited space, which is used as an off-hand controller for click and swipe gestures or other simple interactions. Additionally, these small surfaces are only considered as an off-hand substitute for tangible interface (trackpad / button) on the spectacle frame of smart glasses. As the small surfaces are not advantageous to complicated tasks like text entry, an additional input approach is necessarily vital to fill the gap in the coverage, e.g. speech recognition. 
\paragraph{Considering temporal factor in interaction design}
The timing of switching between multiple input modals is another crucial consideration. Vernier and Nigay \cite{[112]Vernier:2000:FCC:1756227.1756232} proposes a framework to describe five temporal possibilities in input modalities (order, succession, intersection, inclusion, and simultaneity). The key characteristics of the model is to describe the temporal relationship between two or more input approaches. Considering the combination of touch input and freehand interaction, the switching point from touch-based input to mid-air hand gesture can be the manifestation of 3D object. For example, the scenario requires manipulation of virtual 3D object after selecting an application in 2D interface, i.e. succession. Another illustrative example about inclusion can be mingling voice recognition with small-sized touch surface for text entry, as finger-worn device cannot support efficient text entry.
\paragraph{Social acceptance and appealing design}
Among the surveyed papers, social acceptance is regularly included in the evaluation sections. Designing an unobtrusive interaction technique for smart glasses can encourage people to use smart glasses in public area \cite{[106]citeulike:8213624}. As discussed in Section 3, speech recognition has poor social acceptance due to causing disturbance and obtrusion. In contrast, touch-based input has considerably good social acceptance. People nowadays are acceptable to wristbands, rings, and armbands. We can view the touch-sensible external devices as fashionable-traditional gadgets \cite{[58]Shilkrot:2015:DDC:2830539.2828993,[113]Mulling:2015:CHG:2783446.2783612}. Regarding the on-skin interaction, finger-to-palm, thumb-to-finger, and forearm are the most popular touch interfaces \cite{[24]Tung:2015:UGI:2702123.2702214}. However, touch interaction on facial area is uncertain because repetitively touches on the facial area would impact the user's appearance, for instance, removal of make-up or bringing dust on facial area. In addition, one-hand inputs (finger-to-ring and thumb-to-finger interactions) need only subtle interaction and thus avoid awful interactions in public area. As for the freehand interaction, gloves or body-worn cameras are more preferable than head-worn cameras. A study considering social acceptance suggested that the hand gesture should be performed off-face \cite{[81]Hsieh:2016:DWH:2858036.2858436}. The study reported the comments from participants `in-air hand gesture performed in front of the face is weird'. Gloves and body-worn cameras as the form factor might raise the question of why extra device is being worn. We recommend that wearable devices emerge on the market as their outfit designs are considerably attractive. Researchers have to provide aesthetically pleasing appearance to their proposed input devices for higher social acceptance \cite{[40]Weigel:2015:IFS:2702123.2702391}. 
\paragraph{Energy consumption on smart glasses}
Smart glasses have very limited battery life and good utilization of energy can facilitate the everyday use of smart glasses \cite{[111]7273236}. Thus, an additional fundamental factor of energy consumption should be further considered. The energy consumption of the interaction approaches varies from one case to another case. Inputs using external devices or having separate energy provision (e.g. touch-based and glove-based inputs) are preferred choices. In contrast, vision-based approaches using embedded cameras in smart glasses are energy-consuming. It is expected that the energy-consuming issue can be alleviated if multi-modal inputs are appropriately designed. For example, vision-based freehand interactions can be triggered only in some particular scenarios like the interactions with virtual 3D objects are unavoidable, or the cameras will switch on when inertia measurement units inside the finger-worn wearable recognize the forearm movements for hand gestures, and to name but a few.

\section{Conclusions}

In this survey, we studied the smart glasses available on the market, giving a detailed overview of the related literature. We initially presented the research efforts in the field and more specifically in the context of on-device touch input, on-body touch input, hands-free input, and freehand input. We group all these with more abstract terms of touch input and touchless input. We created a classification framework that distinguishes interaction methods for smart glasses, on the basis of their  key characteristics: input modality, form factor,  existence of tactile feedback,  and interaction areas. After that, we categorized and presented the existing research efforts and the interaction challenges on smart glasses. Nevertheless, we see several works have applied multiple input modal to enhance the input capabilities (touch and mid-air gestures), ease-of-use or input accuracy. We believe it is important to further study the trend of multi-modal inputs for smart glasses.  

Although the future of interactions on smart glasses is highly uncertain, the current works, touch and touchless input, give some important clues to the field.  Both the 3D natural hand gestures  and  touch-based gestures  are  important to the smart glasses interaction with the hybrid user interface comprised of 2D and 3D objects.  While there has been significant research on interaction methods using natural hand and touch gestures such as large screen display and touchscreen, very few works (i.e. combining both hand and touch gestures) have been considered in the scenario of augmented reality on mobile devices.  This opens research opportunities for overcoming the hurdle of encumbered interactions with the miniature smart glasses. We propose a potential research direction of creating multi-modal input by combining various input approaches as mentioned in the literature.

\appendix

\bibliographystyle{ACM-Reference-Format}
\bibliography{sample-bibliography}

\end{document}